\documentclass[notitlepage,nofootinbib]{revtex4-1}
\usepackage{amssymb, amsmath, amsopn, color, graphicx}
\usepackage[english]{babel}
\usepackage{float}
\usepackage{bm}
\DeclareMathAlphabet{\matheul}{U}{eus}{m}{n}

\newcommand{\sign}{\mathop{\rm sign}}
\newcommand{\arccsch}{\mathop{\rm arccsch}}

\begin{document}


\title{Formation of dark energy stars}

\author{Philip Beltracchi and Paolo Gondolo}
\email{phipbel@aol.com, paolo.gondolo@utah.edu}
\affiliation{Department of Physics and Astronomy, University of Utah, 115 South 1400 East Suite 201, Salt Lake City, UT 84012-0830}


\renewcommand{\date}[1]{}
\begin{abstract}
\noindent
Dark energy stars are finite size astrophysical objects with an interior equation of state typical of dark energy. Examples are self-gravitating false vacuum bubbles, vacuum nonsingular black holes, and gravastars. We present a time-dependent solution of Einstein's field equations that describes the collapse of a spherical system from an initial state of positive pressure to a final state with a dark energy core. Our solution has no singularities, no event horizons, and does not violate the weak or null energy conditions.
\end{abstract}

\pacs{Valid PACS appear here}
\maketitle

\section{Introduction}

 There are various studies of compact astrophysical objects in the interior of which the energy density $\rho$ and the pressure $p$ obey an equation of state typical of dark energy such as $p=-\rho$. Such objects have been variously named in the literature. We refer to them as ``dark energy stars" for simplicity.
 
 Although dark energy stars could have spacetime singularities, one is commonly interested in dark energy stars that are nonsingular. Buchdahl's theorem \cite{PhysRev.116.1027} precludes the existence of nonsingular compact objects with radius smaller than $9/8$ the Schwarzschild radius under the assumptions of spherical symmetry, isotropic stress, and nonnegative trace of the energy momentum tensor. Compact nonsingular dark energy stars are possible because the nonnegative trace condition does not apply for $p < - \rho/3$ dark energy (compact objects supported by anisotropy instead have also been studied \cite{1974ApJ...188..657B}).

In the mid 1960s the idea of objects with $p=-\rho$ at their center was put forward by Gliner \cite{gliner1966}. The first concrete solution was the Bardeen spacetime \cite{Bardeen,Borde:1994ai,PhysRevD.55.7615,Zhou:2011aa}, which is a nonsingular, asymptotically flat, spherically symmetric spacetime that may have zero, one, or two event horizons depending on the value of a parameter. The Bardeen stress energy tensor features radial pressure $p_r=-\rho$ everywhere and tangential pressure $p_T\ne p_r$ away from the center.

In the 1980s the gravitational effects of false vacuum bubbles forming in true vacuum, and vice versa, were considered  \cite{PhysRevD.21.3305}. False-vacuum bubbles  were studied as a possibility for wormholes \cite{doi:10.1143/PTP.65.1443} and localized inflation \cite{PhysRevD.35.1747,FARHI1987149} when it was found that the null energy condition imposes that any spherically symmetric false-vacuum bubble that forms in an asymptotically flat space, and grows beyond a certain critical size, must have emerged from an initial singularity \cite{FARHI1987149}. Smaller false vacuum bubbles may arise without initial singularities \cite{FARHI1990417}. There were also attempts to replace the black hole singularity inside the horizon with a Planckian density vacuum bubble and a junction layer \cite{FROLOV1989272,PhysRevD.41.383,0264-9381-5-12-002}.

Starting in the 1990s, compact objects with $p=-\rho$ at their center, called vacuum nonsingular black holes, or lambda black holes \cite{Dymnikova1992,Dymnikova2003,Dymnikova:2000zi,Dymnikova:2001mb}, were studied within the class of ``regular black-hole" solutions, i.e., asymptotically flat spacetimes that, like black holes, possess an event horizon but, unlike black holes, do not have a singularity (see, e.g., Ref. \cite{Ansoldi:2008jw} for a review). Lambda black holes, and similar horizonless objects known as G lumps \cite{DYMNIKOVA2007358,Dymnikova:2001fb,Dymnikova:2015yma}, are similar to the Bardeen spacetime in that they have $p_r=-\rho$ everywhere but anisotropic pressure $p_T \ne p_r$. This allows interpolation between a de Sitter core and a Schwarzschild  exterior without junction layers. Interpolations between de Sitter cores and Reissner-Nordstrom exteriors for charged black holes have also been considered \cite{ANSOLDI2007261}. Compact objects with equations of state $p=w\rho$ where $w\ne-1$ were studied in Refs. \cite{doi:10.1139/cjp-2017-0526} ($w<-1/3$) and \cite{Bilic:2005sn} ($w<-1$). 

In the 2000s, the idea of a finite-volume $p=-\rho$ region was revisited as a method of building a gravitationally stable compact object that does not have  singularities or event horizons. The objects described by Chapline \textit{et al}.  \cite{doi:10.1080/13642810108221981} contain a $p=-\rho$ dark energy core and have a microscopic quantum critical layer in place of an event horizon. These objects are described with the term ``dark energy star" in Ref. \cite{Chapline:2005ph} (our usage of the term is more general).  
The stiff shell gravastar proposed by Mazur and Mottola \cite{mazur2001gravitational,MM2004} features a surface layer made of positive pressure stiff matter joined to the dark energy core and exterior vacuum by junction layers. A simplified version of this shell model with an infinitesimal shell was introduced by Visser and Wiltshire \cite{Visser:2003ge}. Whether a gravastar of the Visser-Wiltshire type with particular surface and interior conditions would collapse, explode, stabilize, or oscillate has been studied in Refs. \cite{1475-7516-2008-06-025,1475-7516-2011-10-013}. Anisotropic gravastars with continuous pressure were examined by Cattoen, Faber, and Visser \cite{cattoen2005gravastars} as a means of eliminating the junction layers. Various kinds of gravastars were found to be stable under small perturbations (e.g., Refs. \cite{Visser:2003ge,chirenti2007tell,eosgravastar,0264-9381-22-21-007}), compatible with charge \cite{0264-9381-22-21-007,horvat2008electrically,Chan:2010se,2013JMPh....4..869B}, and with an exterior cosmological constant \cite{chan2010lambda,2013JMPh....4..869B}. The rotation  and angular momentum of gravastars may lead to instability for some rapidly spinning configurations \cite{Cardoso:2007az}, but other spinning configurations are stable \cite{Chirenti:2008pf}. In 2015, Mazur and Mottola examined the Schwarzschild interior solution below the Buchdahl bound, and they found that in the $R\rightarrow R_S$ limit, it behaved as a thin shell gravastar \cite{0264-9381-32-21-215024}. This $R\rightarrow R_S$ Schwarzschild interior gravastar behaves almost exactly as an extended source for the Kerr metric when slow rotation is added \cite{posada2017slowly}. Since gravastars need not have an event horizon, they could in principle be distinguished from black holes \cite{chirenti2007tell}. Gravitational lensing through gravastars has been studied \cite{PhysRevD.90.104013}. There have even been attempts to interpret LIGO data as horizonless compact objects \cite{PhysRevD.94.084016,PhysRevLett.116.171101}.

In this paper, we show that there are time-dependent solutions of Einstein's equations that start with nonnegative pressure $p \ge 0$ everywhere, end in a dark energy star with $p=-\rho$ in a finite central core, have no singularities or junction layers, and do not violate the weak or null energy conditions at any time. We denote the medium composing the system as ``matter" rather than ``fluid" because it involves anisotropic stress. 

We finally remark that in this paper we consider asymptotically flat spacetimes rather than an exterior cosmological constant.

\section{Time varying Spherically symmetric systems}
We write Einstein's field equations for a general time-dependent spherically symmetric system in the form of a force equation and a continuity equation. Spherical symmetry allows the metric to be written in terms of two functions $\Phi(t,r)$ and $m(t,r)$ of the time and radial coordinates $t$ and $r$,
\begin{align}
ds^2 = - e^{2\Phi(t,r)} \, dt^2 + \frac{dr^2}{1-\frac{2Gm(t,r)}{r}} + r^2 \, d\theta^2 + r^2 \, \sin^2\theta \, d\phi^2 .
\end{align}
The corresponding stress-energy tensor $T_{\mu \nu}(t,r)$ may be simplified with tetrads to a local Lorentz frame,
\begin{align}
e^\mu_{\hat{\mu}}e^\nu_{\hat{\nu}}T_{\mu \nu}=T_{\hat{\mu}\hat{\nu}} = \, \begin{pmatrix}
\rho & -S_r & 0 & 0 \\
-S_r & p_r & 0 & 0 \\
0 & 0 & p_T & 0 \\
0 & 0 & 0 & p_T 
\end{pmatrix}
,
\label{eq:T}
\end{align}
\begin{align}
e^\mu_{\hat{\mu}}e^\nu_{\hat{\nu}}g_{\mu \nu}=\eta_{\hat{\mu}\hat{\nu}}=\mathop{\rm diag}(-1,1,1,1), && e^\mu_{\hat{\mu}}=\left(
\begin{array}{cccc}
 e^{-\Phi(t,r)} & 0 & 0 & 0 \\
 0 & \sqrt{1-\frac{2 G m(t,r)}{r}} & 0 & 0 \\
 0 & 0 & \frac{1}{r} & 0 \\
 0 & 0 & 0 & \frac{1}{r\sin\theta} \\
\end{array}
\right).
\label{tetrad}
\end{align}
Here the $T_{\hat{\mu}\hat{\nu}}$ are components of the stress-energy tensor in an inertial frame at rest in the $t,r,\theta,\phi$ coordinate system. More specifically, $\rho=T_{\hat{t}\hat{t}}$ is the energy density, $p_r=T_{\hat{r}\hat{r}}$ and $p_T = T_{\hat{\theta}\hat{\theta}}= T_{\hat{\phi}\hat{\phi}}$ are the radial and transverse stresses, and $S_r = -T_{\hat{r}\hat{t}} = -T_{\hat{t}\hat{r}}$ is the $r$-component of the momentum density, with positive $S_r$ corresponding to the outward flow. We define our system in terms of the matter functions $\rho(t,r)$, $p_r(t,r)$, $\Delta(t,r) = p_T(t,r) - p_r(t,r)$, and $S_r(t,r)$, since they are more closely related to the weak energy condition. The function $\Delta(t,r)$ embodies a possible anisotropic stress. Although $p_r$ and $p_T$ should properly be referred to as radial and tangential stress, we follow the existing literature and call them radial and tangential pressure.


Einstein's equations  become
\begin{align}
 \frac{\partial m}{\partial r} & = 4\pi r^2 \rho ,
\label{eq:Einstein-rho}
\\[1ex]
\frac{\partial\Phi}{\partial r} & = \frac{G(m+4\pi r^3 p_r)}{r^2 \left(1-\frac{2Gm}{r}\right)} ,
\label{eq:Einstein-pr}
\\[1ex]
\frac{\partial m}{\partial \tau} & = -4\pi r^2\sqrt{1-\frac{2Gm}{r}} S_r ,
\label{eq:sr1}
\\[1ex]
- \frac{\partial p_r}{\partial r}-\frac{G \left( m+4\pi r^3 p_r \right) \left( \rho+p_r \right)}{r^2 \left( 1 - \frac{2 G m}{r} \right) }+\frac{2\Delta}{r}&=\sqrt{1-\frac{2Gm}{r}} \frac{\partial}{\partial\tau} \left( \frac{S_r}{1-\frac{2Gm}{r}} \right).
   \label{Eq:Force}
\end{align}
Here we have introduced a new time variable $\tau$ defined so that
\begin{equation}
    e^{-\Phi(t,r)} \frac{\partial}{\partial t} = \frac{\partial}{\partial \tau}.
    \label{taudef}
\end{equation}

Equation~(\ref{eq:sr1}) and Eq.~(\ref{eq:Einstein-rho}) can be rearranged into a continuity equation for the energy density $\rho$ and energy flux $S_r$,
\begin{equation}
    \frac{1}{r^2}\frac{\partial}{\partial r}\left(r^2\sqrt{1-\frac{2Gm}{r}}S_r \right)+\frac{\partial \rho}{\partial \tau}= 0 .
\label{eq:Continutiy}
\end{equation}

Equation~(\ref{Eq:Force}) resembles Newton's second law for the force density, where we can identify the terms on the left as the pressure gradient force, gravitational force, and anisotropy force, respectively, while the term on the right-hand side embodies the rate of change of momentum. Notice that the anisotropy force $2\Delta / r$ is a nonrelativistic force \cite{1974ApJ...188..657B,cattoen2005gravastars} coming from the spatial divergence of an anisotropic stress tensor.

 Equation~(\ref{eq:Einstein-rho}) and Eq.~(\ref{eq:Einstein-pr}) are easily solved for $m(t,r)$ and $\Phi(t,r)$ with boundary conditions $m(t,r)=0$ at $r=0$ and $\Phi(t,r)=0$ at $r\to\infty$, for any $t$,
\begin{align}
m(t,r) & = \int_0^r \, \rho(t,r) \, 4 \pi r^2 \, dr ,
\label{eq:m(t,r)}
\\
\Phi(t,r) & = - \int_r^\infty  \frac{G \left( m + 4 \pi r^3 p_r \right) }{ r^2 \left( 1 - \frac{2Gm}{r} \right) } \, dr .
\label{eq:Phi(t,r)}
\end{align}
We can also rearrange Eq.~(\ref{eq:sr1}) and Eq.~(\ref{Eq:Force}) to solve for $S_r$ and $\Delta$
\begin{align}
    S_r & = -\frac{1}{\sqrt{1-\frac{2Gm}{r}}}  \frac{1}{4\pi r^2}  \frac{\partial m}{\partial \tau},
\label{eq:Einstein-Sr}
\\
\Delta & =\frac{r}{2}\Bigg[\frac{\partial p_r}{\partial r}+\frac{G \left( m+4\pi r^3 p_r \right) \left( \rho+p_r \right)}{r^2 \left( 1 - \frac{2 G m}{r} \right) }+\sqrt{1-\frac{2Gm}{r}} \frac{\partial}{\partial\tau} \left( \frac{S_r}{1-\frac{2Gm}{r}} \right)\Bigg].
\label{eq:Einstein-Delta}
\end{align}
Notice that if one would impose static anisotropic conditions, from Eq.~(\ref{Eq:Force}) the gradient of the pressure would have to satisfy
\begin{align}
\frac{dp_r}{dr} = - \frac{G \left( m+4\pi r^3 p_r \right) \left( \rho+p_r \right)}{r^2 \left( 1 - \frac{2 G m}{r} \right) } + \frac{2\Delta}{r},
\end{align}
which is the Tolman--Oppenheimer--Volkoff equation with an extra term due to the anisotropy. 

Thin shells of matter manifest as Dirac delta functions or derivatives of Dirac delta functions in the components of the stress-energy tensor $T_{\mu\nu}$, which in our case are the functions $\rho$, $p_r$, $S_r$, and $\Delta$. There are no thin shells in $\rho$ or $S_r$ if the function $m(t,r)$ is continuous, as can seen from Eqs.~(\ref{eq:Einstein-rho}) and~(\ref{eq:Einstein-Sr}). There are no thin shells in $p_r$ if the function $\Phi(t,r)$ is continuous, as can be seen from Eq.~(\ref{eq:Einstein-pr}). There are no thin shells in $\Delta$ if the function $p_r(t,r)$ is continuous and the function $m(t,r)$ has continuous first derivatives, as can be seen by rewriting the last term in Eq.~(\ref{eq:Einstein-Delta}) as
\begin{align}
   \frac{r}{2} \sqrt{1-\frac{2Gm}{r}} e^{-\Phi} \frac{\partial}{\partial t} \left( \frac{S_r}{1-\frac{2Gm}{r}} \right)
=
   \frac{1}{8\pi(r-2Gm)} e^{-2\Phi} \left[ \frac{\partial \Phi}{\partial t} \frac{\partial m}{\partial t}
   - \frac{\partial^2 m}{\partial t^2}
   - \frac{3G}{r-2Gm} \left(\frac{\partial m}{\partial t}\right)^2 \right]
\end{align}
and
\begin{align}
\frac{\partial \Phi}{\partial t} = 
- G \int_r^\infty \left[ \frac{1+8\pi G r^2 p_r}{(r-2Gm)^2} \frac{\partial m}{\partial t} + \frac{4\pi r^2}{r-2Gm} \frac{\partial p_r}{\partial t} \right]\, dr .
\end{align}

We conclude this section by mentioning 
that Birkhoff's form of his theorem \cite{Birchoff} states that if a spherically symmetric system is surrounded by empty space, i.e., $T_{\mu \nu}=0$ beyond a certain radius $R$, its total mass $M$ is constant. This can be seen from our work as follows: the total mass $M=m(t,R)$, where we take the limit $r\rightarrow R$ with $r>R$; the condition $T_{\mu \nu}=0$ at $r=R$ requires $S_r(t,R)=0$, and from Eq.~(\ref{eq:Einstein-Sr}), $\partial m(t,R)/\partial t=0$, which is $dM/dt=0$. 
\section{Energy conditions}
We specify the null and weak energy conditions here for the stress-energy tensor in Eq.~(\ref{eq:T})  (information on energy conditions can be found in Ref. \cite{Curiel2017}). 

One common method for considering the energy conditions is to put the stress-energy tensor into one of the canonical types (see, e.g., Ref. \cite{Hawking:1973uf}). The stress-energy tensor in Eq.~(\ref{eq:T}) is either type I or type IV. It is type I when $(\rho+p_r)^2\ge 4 S_r^2$, and it is type IV when $(\rho+p_r)^2< 4 S_r^2$. If it is type IV, the weak energy condition cannot be satisfied \cite{Hawking:1973uf}, and so we do not consider it. If it is type I, then its canonical form is
\begin{equation}
    T_{\mu \nu}=\rho^0 u_\mu u_\nu+p_r^0 \chi_\mu \chi_\nu+p_T^0(g_{\mu \nu}+u_\mu u_\nu-\chi_\mu \chi_\nu),
\end{equation}
where $u^\mu$ is the four-velocity of the matter, $\chi^\mu$ is a unit vector in the radial direction, and $\rho^0$, $p_r^0$, $p_T^0$ are the proper density and proper principal pressures of the matter (in the matter rest frame)
\begin{equation}
    \rho^0=\frac{\rho-p_r+y}{2},\quad p_r^0=\frac{p_r-\rho+y}{2},\quad p_T^0=p_T,\quad y=\sqrt{(\rho+p_r)^2-4 S_r^2}.
    \label{restframe}
\end{equation}
In the local Lorentz frame defined by the tetrad in Eq.~(\ref{tetrad}), the components of the four-vectors $u^\mu$ and $\chi^\mu$ are
\begin{equation}
    u^{\hat{\mu}}=\big(\frac{1}{\sqrt{1-v_r^2}},\frac{v_r}{\sqrt{1-v_r^2}},0,0\big),\qquad \chi^{\hat{\mu}}=\big(\frac{v_r}{\sqrt{1-v_r^2}},\frac{1}{\sqrt{1-v_r^2}},0,0\big).
\end{equation}
Here $v_r$ is the radial velocity of the matter (negative for infall)
\begin{equation}
    v=\sign(S_r)\sqrt{\frac{\rho+p_r-y}{\rho+p_r+y}}.
\end{equation}
The weak energy condition in the matter rest frame then reads
\begin{equation}
    \rho^0\ge0,\qquad \rho^0+p_r^0\ge0,\qquad \rho^0+p_T^0\ge0.
    \label{wecrest}
\end{equation}
Since we later assign the functions $\rho(t,r)$ and $p_r(t,r)$ instead of $\rho^0$ and $p_r^0$, the weak energy condition in Eq.~(\ref{wecrest}) assumes a complicated form because of the presence of the square root in $y$. We therefore find simpler expressions for the weak energy condition directly rather than from the canonical form.

The weak energy condition (WEC) is $T_{\mu\nu} k^\mu k^\nu \ge 0$ for all timelike (and in the limiting case lightlike) vectors $k^\mu$. The quantity $T_{\mu\nu} k^\mu k^\nu/(-k^2)$ is the energy density measured by an observer with four velocity $k^\mu/\sqrt{-k^2}$. The components of $k^\mu$ in a local Lorentz frame can be parametrized as
\begin{align}
k^{\hat{\mu}} = (k^{\hat{t}},k^{\hat{r}} ,k^{\hat{\theta}} ,k^{\hat{\phi}} ) = E \, (1, \beta  \cos\alpha, \beta  \sin\alpha \cos\varphi, \beta   \sin\alpha \sin\varphi) , 
\end{align}
with $0\le\beta \le 1$, $0\le\alpha\le\pi$, and $0\le\phi<2\pi$.
Then the weak energy condition becomes
\begin{equation}
 \rho + \beta^2 p_r \cos^2\alpha + \beta^2 p_T \sin^2\alpha - 2 \beta S_r \cos\alpha \ge 0
 \label{eq:weakcondition}
\end{equation}
for all $\alpha$ and for $0\le \beta \le 1$. Depending on the parameters $\rho, p_r, p_T, S_r$, the minimum must either be on the boundary of the region $0\le\beta\le1$, $-1\le\cos{\alpha}\le1$ or be a local minimum inside the region. The weak energy condition amounts to the following inequalities:
 \begin{align}
 &\text{ if $p_r - | S_r| \ge 0$ and $p_T \ge 0$,}&& \text{the WEC is }\rho-\frac{S_r^2}{p_r}\ge0 , \label{eq:wec3} \\
&\text{ if $p_r - | S_r| \le 0$ and $p_T \ge p_r - | S_r|$,}&& \text{the WEC is } \rho+p_r- 2|S_r|\ge0,  \label{eq:wec2} \\
& \text{ if $p_T \le p_r - | S_r|$ and $p_T \le 0$,}&& \text{the WEC is }\rho+p_T+\frac{S_r^2}{p_T-p_r}\ge0 .  \label{eq:wec4} 
\end{align}
A compact way of writing these inequalities is 
 \begin{align}
     \rho+p_r-W-\frac{S_r^2}{W}\ge0,&& W=\max(p_r,|S_r|,p_r-p_T) .
 \end{align}
 Setting $S_r=0$ we recover the well-known static case 
\begin{align}
\rho \ge 0, \quad \rho+p_r \ge 0, \quad \rho+p_T \ge 0 \qquad \text{(for $S_r=0$).} \label{eq:wecstat}
\end{align}
Note that Eq.~(\ref{eq:wec2}) enforces reality of $\rho^0$ and $p_r^0$ and forces the stress-energy tensor to be type I.
In other words, the rest frame for a matter element (a frame in which the energy flow vanishes, $S_i=0$, $i=1,2,3$) is given by a boost of velocity $v_r$ from our standard frame. The energy condition $\rho+p_r-2|S_r|\ge0$ implies the reality of $y$ in Eq.~(\ref{restframe}), whereas a violation implies imaginary $y$, complex $v_r$, and nonexistence of a rest frame for the matter element. In other terms, if no inertial frame has $S_i=0$, then in some inertial frame the energy density is negative.

The null energy condition (NEC) is   $T_{\mu\nu} k^\mu k^\nu \ge 0$ for all lightlike vectors $k^\mu$. It similarly becomes
\begin{align}
    &\text{ if $p_T \le p_r - | S_r|$,}&& \text{the NEC is } \rho+p_T+\frac{S_r^2}{p_T-p_r}\ge0  ,
    \\
    & \text{ if $p_T > p_r - | S_r|$,}&& \text{the NEC is  }\rho+p_r- 2|S_r|\ge0 . 
\end{align}
The weak energy condition implies the null energy condition.

For completeness, we recall that the strong energy condition (SEC) is $(T_{\mu\nu}-\tfrac{1}{2} T^\lambda_{\hspace{0.5 em} \lambda} g_{\mu\nu}) k^\mu k^\nu \ge 0$ for all timelike vectors $k^\mu$, and the dominant energy condition (DEC) is $T_{\mu\nu} k^\mu k^\nu \ge 0$ and $T^\lambda_{\hspace{0.5 em} \mu} T_{\lambda\nu} k^\mu k^\nu \le 0$ for all timelike vectors $k^\mu$.

\section{Pileup models}
We now introduce a class of models for the formation of dark energy stars that describe the collapse of a system from an initial state of positive pressure to a final state with a dark energy core, defined as a central region where $p_r=p_T=-\rho=$constant. We call our class of models pileup models because we build the dark energy core progressively by ``piling up" matter onto its surface.  In pileup models, the energy density at the center increases until it reaches its final value. The pressure at the center initially increases with the density, then decreases until it reaches a value of $p=-\rho$, at which point the dark energy core is formed. After this, the dark energy core expands outward. We call an object where the density at the center has not yet reached its maximum value a precursor and an object with $p=-\rho$ at the center a dark energy star. We call an intermediate stage, if present, the transition.

We build our dark energy core by adding matter to its surface without changing its density so as to satisfy the WEC (\ref{eq:wec2}). Indeed, in the dark energy core, where $p_r=p_T=-\rho$, Eq.~(\ref{eq:wec2}) implies that $S_r$ must be 0  to avoid violating the weak energy condition. As a consequence the mass within any sphere contained in the dark energy core must be constant in $t$, and since this is true of every sphere in the dark energy core, the density must be constant in $t$ as well.

Spatially, the precursor is a ``normal" object with positive pressure and pressure gradient force pointing outwards opposing the force of gravity. The dark energy star has three zones: an innermost dark energy core where the density and pressure are constant in both space and time, an outermost normal zone where the pressure gradient force points outwards, and an intermediate region we call the inversion zone where the pressure gradient force points inwards. The inversion zone is necessary for the radial pressure to be a continuous function of the radius while being negative in the dark energy core and positive in the normal zone.

In this paper we  specify the time and radial dependence in $\rho$ and $p_r$ in a way that avoids singularities and event horizons and does not violate the weak (and therefore null) energy condition. Then we derive $S_r$ from Eq.~(\ref{eq:Einstein-Sr}), and $\Delta$, and hence $p_T$, from Eq.~(\ref{eq:Einstein-Delta}). We use these derived functions to check the validity of the  energy conditions Eqs.~(\ref{eq:wec3})-(\ref{eq:wec4}). The metric functions $m$ and $\Phi$ follow from Eqs.~(\ref{eq:m(t,r)}) and (\ref{eq:Phi(t,r)}). The contributions to $\Phi$ from the dark energy core and exterior Schwarzschild vacuum are closed form, but the contribution from the inversion and normal zones in general needs to be evaluated numerically.
\subsection{Example of pileup model}
Here, we present a parametrization of $\rho$ and $p_r$ for the formation of a dark energy star with total mass $M$. As an aid in avoiding event horizons and singularities while maintaining the WEC, we introduce an evolution parameter $f$ that increases monotonically during the collapse. In subsection B, we relate $f$ to the time $t$. At $f=0$, the density at the center reaches the value of the density in the dark energy core, and at $f=f_D$, the dark energy core is formed.

The density function $\rho(t,r)$ has a great deal of freedom when using this framework, but there are still some restrictions. To make the radius $R(t)$ of a dark energy star similar to that of a black hole, we set the density in the core as the Schwarzschild density $\rho_S=3M/(4\pi R_S^3)$, where $R_S=2GM$ is the Schwarzschild radius.
Birkhoff's theorem requires
\begin{align}
\int_0^{R(t)} 4 \pi r^2 \rho(t,r) \, dr=M ,
\end{align}
where $M$ is a constant. For simplicity, we use straight lines in the parametrization of $\rho$. In the precursor stage we set
\begin{equation}
    \rho(\text{precursor}, f<0)=\begin{cases}
 0, & x\geq s, \\
 \frac{4 \rho_S}{s^4}(s-x), &x<s,
\end{cases}
\label{rhopre}
\end{equation}
where $x=r/R_S$. The parameter $s$ gives the star radius through $R=s R_S$. For the transition stage and dark energy star stage, we set
\begin{equation}
    \rho(\text{transition and dark energy star}, f\ge0)=\begin{cases}
 0, & x\geq s, \\
\rho_S \frac{s-x}{s-f},&f<x<s,\\
\rho_S, &0\leq x\leq f.
\end{cases}
\label{rhopost}
\end{equation}
  For positive $f$, the radius of the constant density plateau is $R_p=f R_S$. Demanding that $M$ be constant at all times requires the following relationship between $f$ and $s$:
\begin{align}
(s-f) (s^3+fs^2+f^2s+f^3-4)=0.
\end{align}
The only real solution besides the trivial $s=f$ is
\begin{align}
s = - \frac{f}{3} \left[ 1 + 2 \sqrt{2} \sinh\left( \frac{1}{3} \arccsch \frac{\sqrt{2} f^3}{5f^3-27} \right) \right] .
\end{align}
When $f=1$, we have $s=1$, the radius of the object equals the Schwarzschild radius, and an event horizon at $r=R_S$ appears in the exterior Schwarzschild metric. When $f=0$, we have $s=4^{1/3}$ and both Eqs.~(\ref{rhopre}) and (\ref{rhopost}) for $x<s$ reduce to $\rho=\rho_S(1-x/s)$, showing that the density is continuous across $f=0$.

\begin{figure}[H]
    \centering
       \includegraphics[width=9cm]{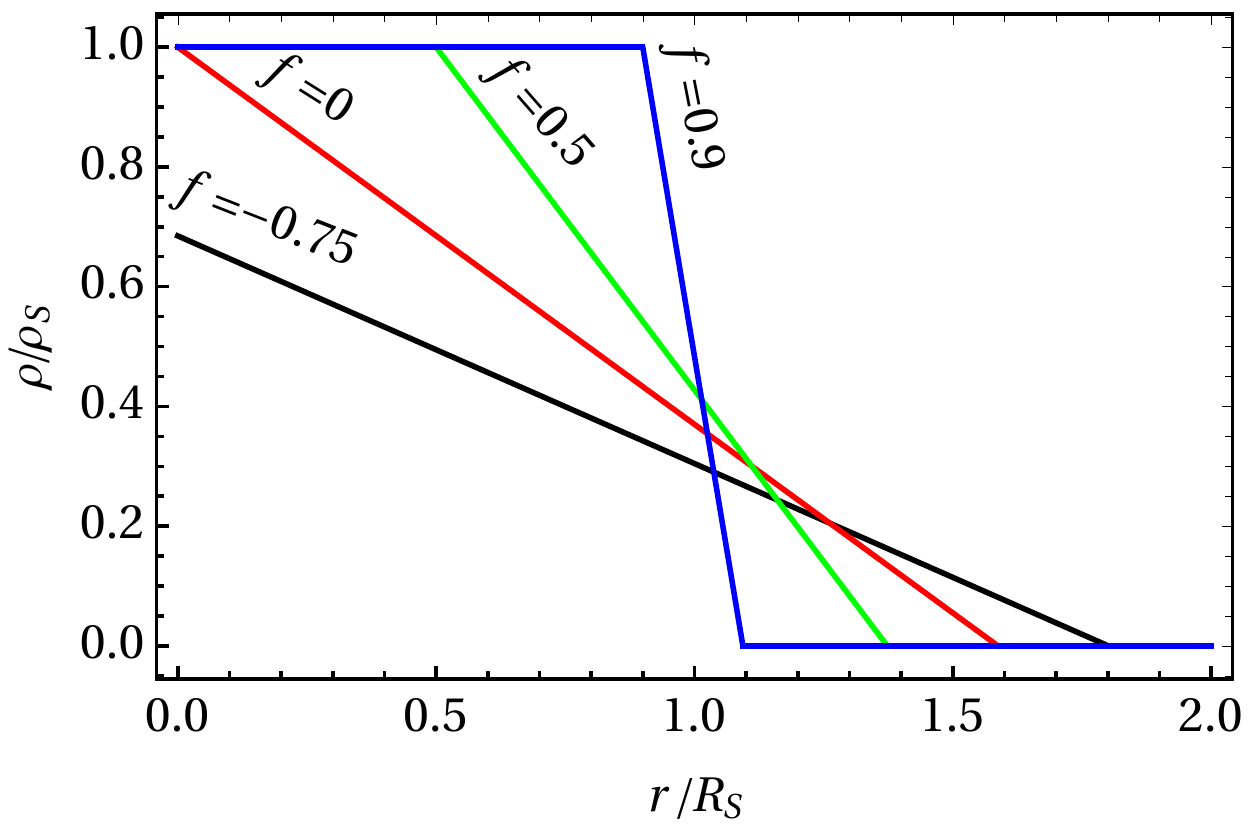}
        \caption{ Density profiles at various stages of collapse in the example pileup model. The density is in units of the core density $\rho_S$, and the radius is in units of the Schwarzschild radius $R_S$. The evolution parameter $f$ describes the stage of collapse. The density profiles are chosen such that the density has a flat region to contain the core, a linearly decreasing normal zone for simplicity, and constant total mass. }
     \label{rainbow den}
\end{figure}

Figure \ref{rainbow den} shows the density profiles Eqs.~(\ref{rhopre}), (\ref{rhopost}) at various stages of collapse, in the precursor stage ($f=-0.75$), at the beginning of the transition stage ($f=0$), and during the dark energy star stage ($f=0.5$,$f=0.9$). The density of the plateau remains constant and the radius of the plateau increases with time.

For the radial pressure function $p_r(t,r)$ we require: (i) continuity at the surface $p_r(r\ge R)=0$, (ii) $p_r=-\rho$ within the core, (iii) $p_r\ge-\rho$ everywhere not to violate the WEC, (iv) $p_r\ge 0$ in the precursor stage, and finally (v) $p_r$ to be a continuously differentiable function in $r$ to avoid singularities and for greater regularity in $p_T$.\\

We set
\begin{equation}
  p_r(\text{precursor}, f<0)=   
\begin{cases}
\rho_S  \frac{4 a}{s^3} \cos ^4\! \left(\frac{\pi  x}{2 s}\right), &  x\leq s, \\
 0, & x>s;
\end{cases}
  \label{eqpr1}
\end{equation}
\begin{equation}
  p_r(\text{transition}, 0\leq f<f_D)=   
\begin{cases}
\rho_S a+\rho_S(1+a)\frac{f}{f_D} \Big[\Psi \! \left( \frac{f-x}{f} \right)-1\Big],& 0\leq x<f,\\
 \rho_S a \cos ^4 \! \left(\frac{\pi  (x-f)}{2 (s-f)}\right), & f<x\leq s,\\
 0, & x>s;
\end{cases}
  \label{eqpr2}
\end{equation}
\begin{equation}
  p_r(\text{dark energy star}, f\ge f_D)=   
\begin{cases}
-\rho_S, &x<f-f_D,\\
\rho_S\Big[-1+(1+a) \Psi \! \left(\frac{f-x}{f_D}\right)\Big], & f-f_D\leq x \leq f,\\
 \rho_S a \cos ^4\! \left(\frac{\pi  (x-f)}{2 (s-f)}\right), & f<x\leq s,\\
 0, & x>s.
\end{cases}
  \label{eqpr3}
\end{equation}
Here, $\Psi(\xi)$ is the following function that smoothly interpolates between 1 and 0
\begin{equation}
     \Psi(\xi)= \begin{cases}
     1 & \xi<0\\
    e^{1-\frac{1}{1-\xi^2}}, & 0\leq \xi \le1,\\
    0, & \xi\ge1.
    \end{cases}
\end{equation}

 The parameter $a$ is the ratio of the radial pressure to the density at the outer edge of the inversion zone. For $a$, we choose $a=0.315$ such that the maximum value of $p_r/\rho$ is equal to $1/3$, which is the value for radiation. For the parameter $f_D$, we use $f_D=0.25$. In this example, the inversion zone and dark energy core are within the constant density plateau. This is necessary for the dark energy core, but the inversion zone can in principle extend outside the plateau.
 
\begin{figure}[H]
    \centering
        \includegraphics[width=9cm]{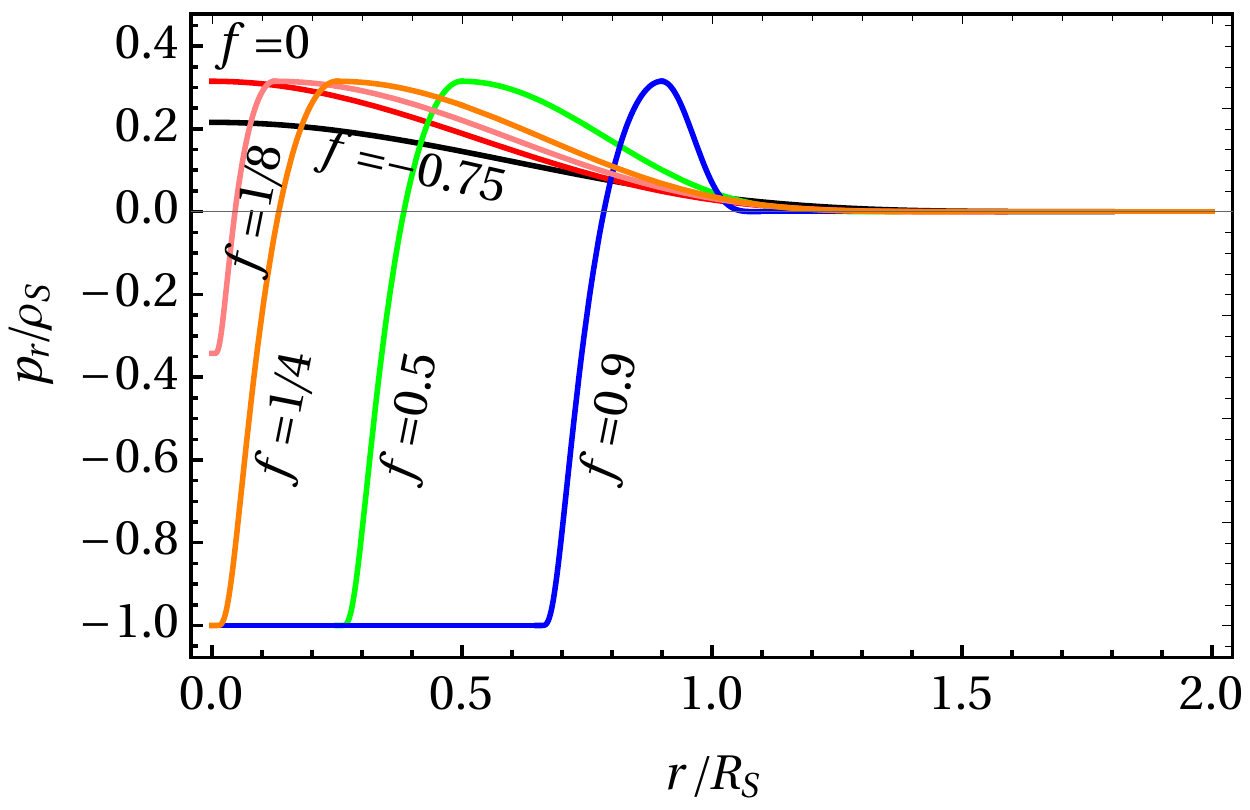}
        \caption{ Radial pressure for the same stages of collapse as in Fig. \ref{rainbow den}, plus two extra contours showing the formation of the inversion zone.}
    \label{rainbow pr}
\end{figure}

 Figure \ref{rainbow pr} shows the radial pressure profiles Eqs.~(\ref{eqpr1})-(\ref{eqpr3}) for the same stages of collapse as in Fig. \ref{rainbow den} plus two extra stages ($f=1/4$, $f=1/8$) to show the formation of the inversion zone.
\subsection{End states and choice of $f(t)$}
\begin{figure}[H]
    \centering
    \includegraphics[width=8cm]{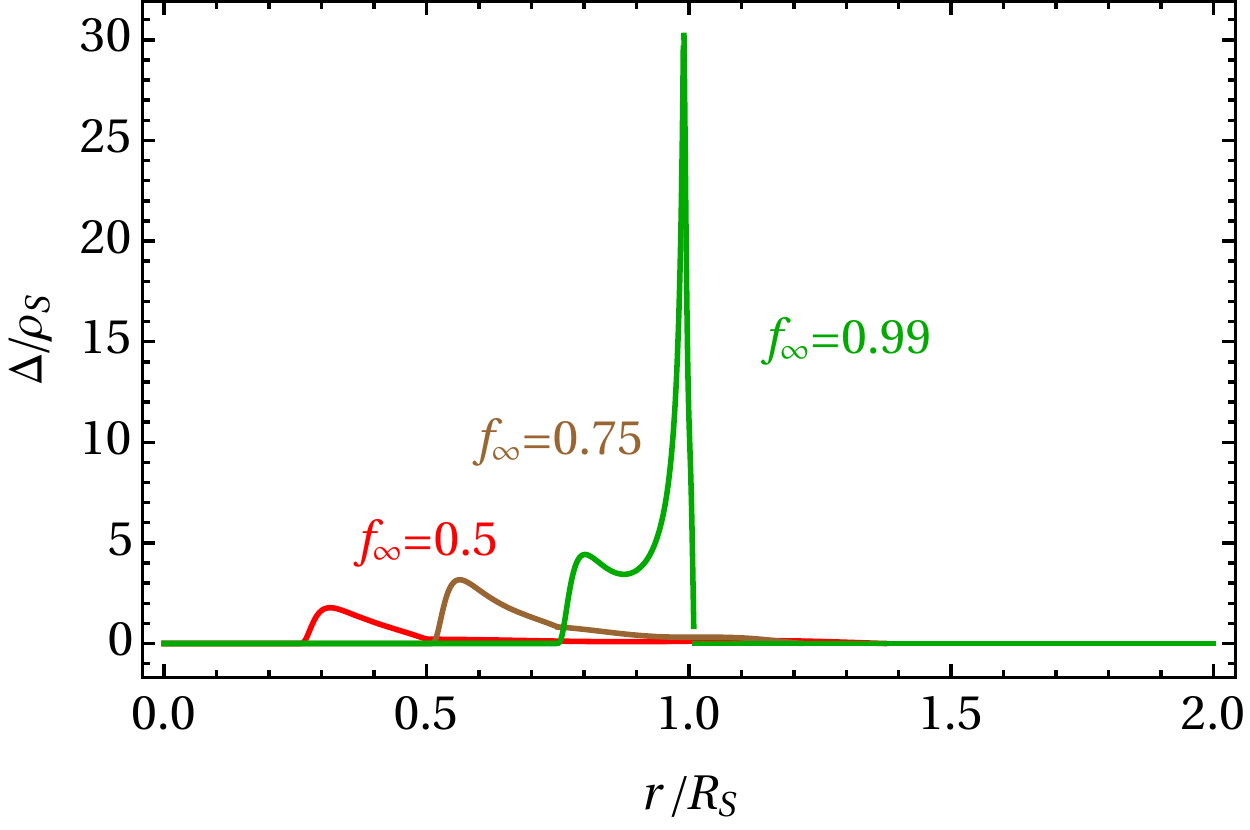}
    \caption{Anisotropies from various possible end states. The lines are $f_\infty=0.5$ (red), $f_\infty=0.75$ (brown), and $f_\infty=0.99$ (green). Note the pronounced positive anisotropy in the inversion zone and near $r=R_S$ as $R\rightarrow R_S$ (visible in green curve). Despite the pressure gradient term in the anisotropy becoming large and negative, the total anisotropy in the normal zone remains positive because the positive gravitational term also increases.}
    \label{endstates}
\end{figure}

   A configuration that has reached a value $f=f_\infty$, and is no longer changing in $t$, we denote as an ``end state." By the chain rule $\partial m/\partial t=(\partial m/\partial f )(\partial f/\partial t)$ and Eq.~(\ref{eq:Einstein-Sr}), end states have $S_r=0$. The threshold for event horizon formation is $f=s=1$. We may prevent event horizons by specifying $f_\infty<1$. Still, the WEC may be violated at a particular $f_\infty<1$ due to a large negative tangential pressure $p_T$ arising from the $\partial p_r/\partial r$ term of Eq.~(\ref{eq:Einstein-Delta}) becoming large and negative as $f$ and $s$ approach 1. We may prevent such a violation of the WEC either by choosing a suitable radial profile for $p_r$ [such that a large positive gravity term in Eq.~(\ref{eq:Einstein-Delta}) cancels the large negative pressure gradient term  in Eq.~(\ref{eq:Einstein-Delta})] or by specifying a more restrictive $f_\infty$ (reducing the pressure gradient term by having an object with a large radius $R$). We do the former, exploiting the fact that the gravity term in the anisotropy Eq.~(\ref{eq:Einstein-Delta}) becomes large and positive if $f_\infty$ is close to 1. Figure \ref{endstates} shows the anisotropy $\Delta$ for end states at $f_\infty=0.5,0.75,0.99$ for the radial pressure and density as defined in Subsection A. 
  It is also illustrative to examine the end states in terms of forces using Eq.~(\ref{Eq:Force}). We introduce the following notation for the pressure gradient, gravitational, and  anisotropy force densities (negative forces point inwards):
  \begin{align}
      F_p &=-\frac{\partial p_r}{\partial r},\\
      F_G &=-\frac{G \left( m+4\pi r^3 p_r \right) \left( \rho+p_r \right)}{r^2 \left( 1 - \frac{2 G m}{r} \right) },\\
      F_\Delta &=\frac{2\Delta}{r}.
  \end{align}
 \begin{figure}[H]
    \centering
    \includegraphics[width=5.8cm]{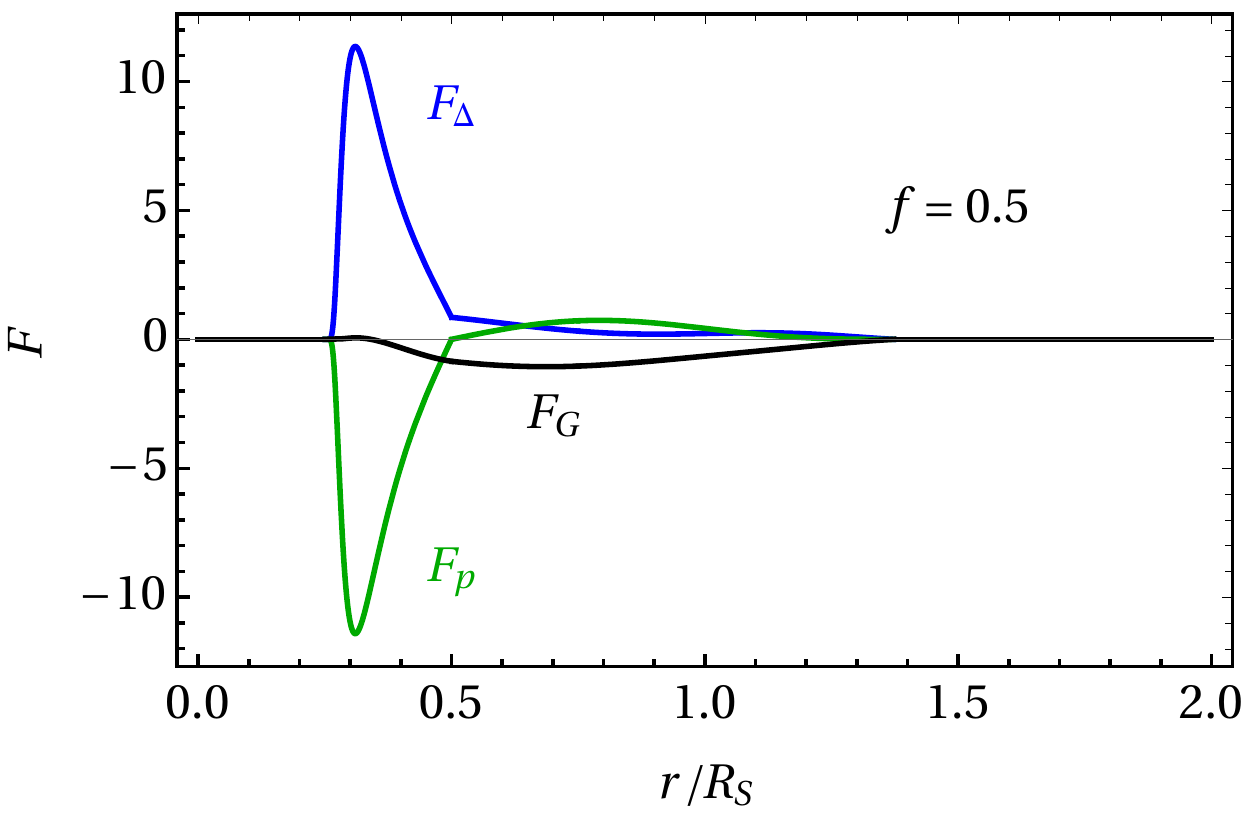}
    \includegraphics[width=5.8cm]{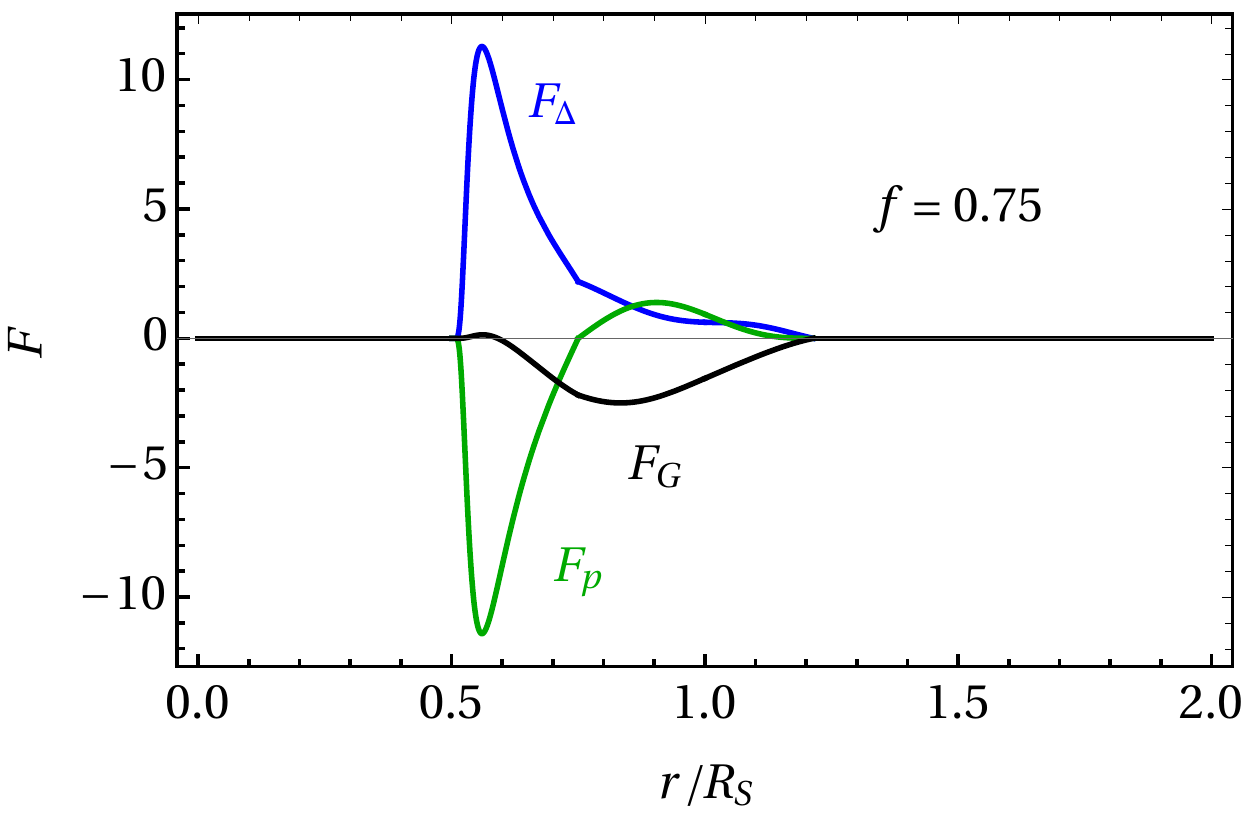}
    \includegraphics[width=5.8cm]{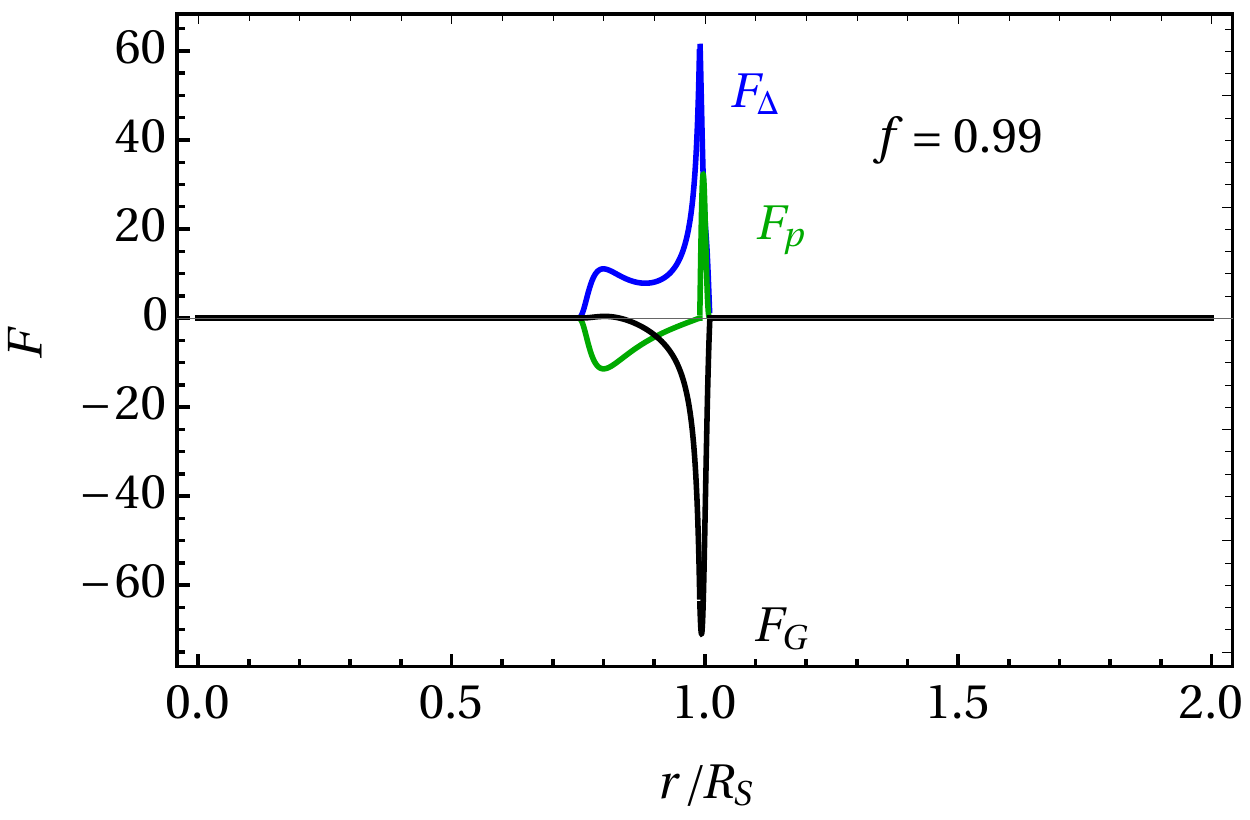}
    \caption{Force densities for the same end state configurations shown in Fig. \ref{endstates}. $F_p$ (green line), $F_G$ (black line), and $F_\Delta$ (blue line) are the pressure gradient, gravitational, and anisotropy force densities, respectively. Negative forces point inwards. The small $r$ region, where the three force densities are zero, is the dark energy core. In the inversion zone the anisotropy force $F_\Delta$  largely cancels the inward pressure gradient force $F_p$, although if one looks closely one can see that the gravity force pushes outward in the lower inversion zone. As $f_\infty$ gets close to one, the gravity force $F_G$ becomes much stronger, and the other forces increase to compensate it. }
    \label{endforces}
\end{figure}  
  Since the changing momentum terms on the right side of Eq.~(\ref{Eq:Force}) are zero for end states, examining the forces in end states gives an idea of how the object is supported, see Fig. \ref{endforces}. We see that the anisotropy force is in fact the only outward force in the part of the inversion zone with positive pressure, in line with the fact that continuous pressure gravastars require anisotropy \cite{cattoen2005gravastars}. If the pressure in the inversion zone is within a negative range, specifically $-\rho_S<p_r<-\rho_S/3$, the gravitational force pushes outwards. Repulsive gravity and anisotropy are unavoidable consequences of interpolating continuously between a dark energy core and a region with positive pressure.
  
We choose the relation $f=f(t)$ between the evolution parameter $f$ and the time $t$ in a way to avoid singularities, event horizons, and/or violation of the weak energy condition. For our example we use the function 
  \begin{equation}
     f(t)=\begin{cases}
 -f_\infty, & t<-\frac{t_C}{2}, \\
  \frac{f_\infty}{4} \Big[15\frac{t}{t_C} -40 \left(\frac{t}{t_C}\right)^3+48 \left(\frac{t}{t_C}\right)^5\Big], & -\frac{t_C}{2}\le t \le \frac{t_C}{2}, \\
  f_\infty, &  t>\frac{t_C}{2}. 
\end{cases}
    \label{ftau}
\end{equation}
Here, $t_C$ is the total collapse time, which we set as $t_C=30 R_S$. Setting $f_\infty=0.9$ avoids problems with the WEC because $p_T$ in the normal zone remains positive, avoids event horizons because $f_\infty<1$, and allows the radius of the dark energy star to become smaller than the Buchdahl bound of $(9/8) R_S$ in that $R_\infty=s_\infty R_S=1.094R_S$. 

\begin{figure}[H]
    \centering
    \includegraphics[width=5.8cm]{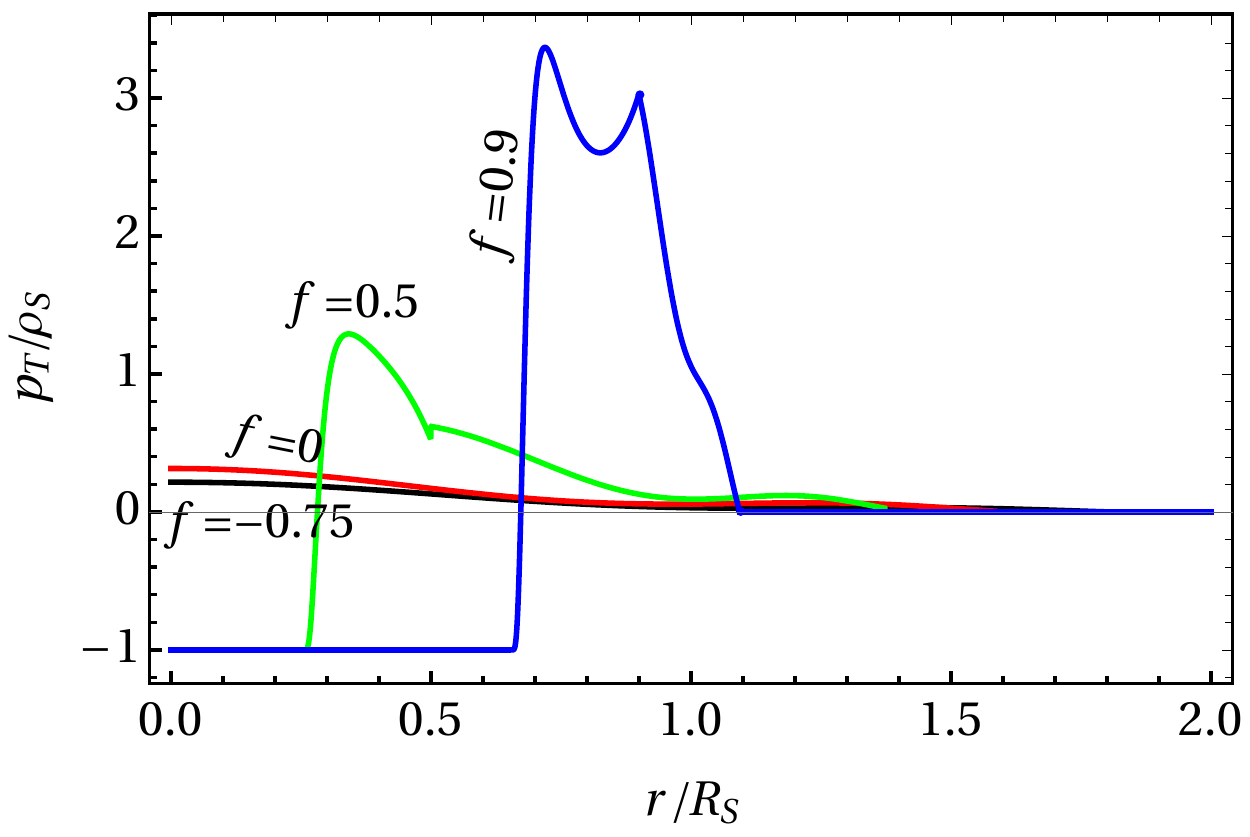}
    \includegraphics[width=5.67cm]{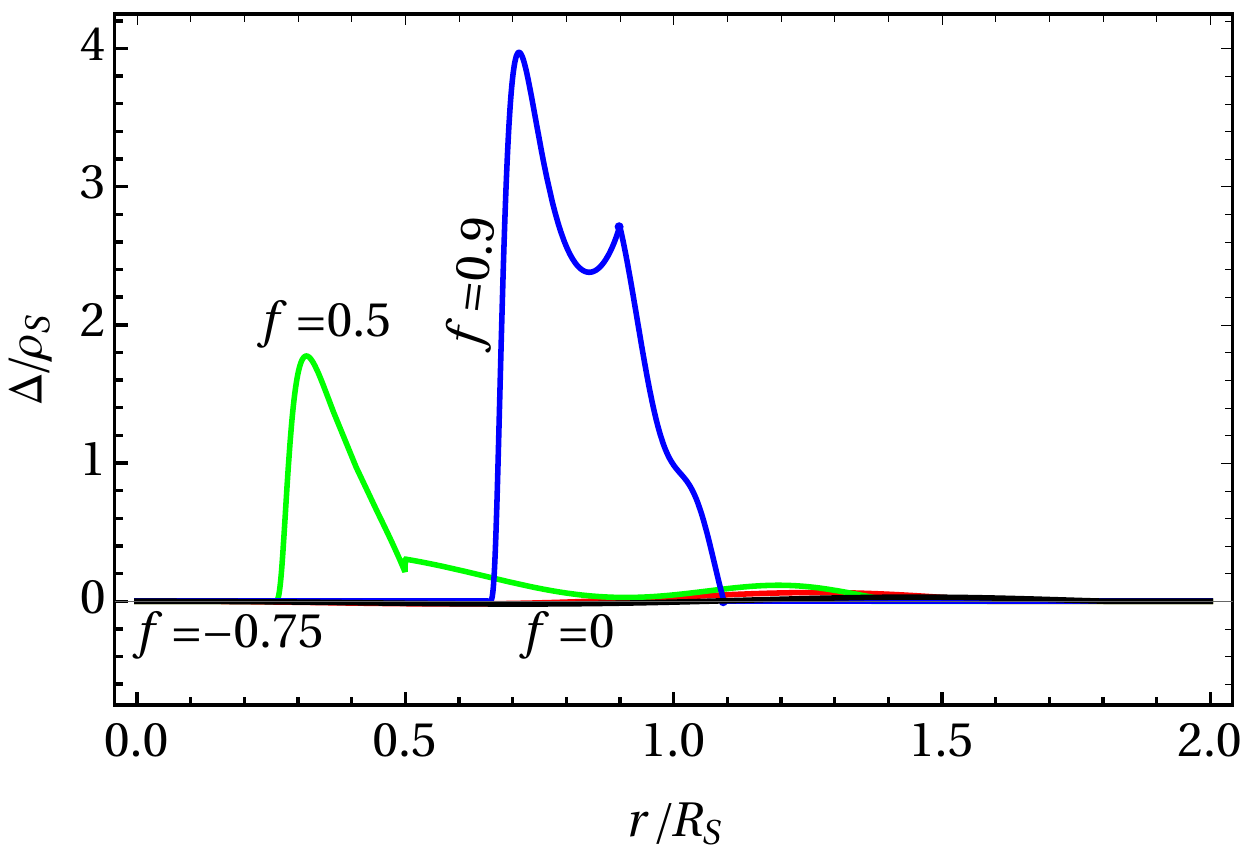}
    \includegraphics[width=5.94cm]{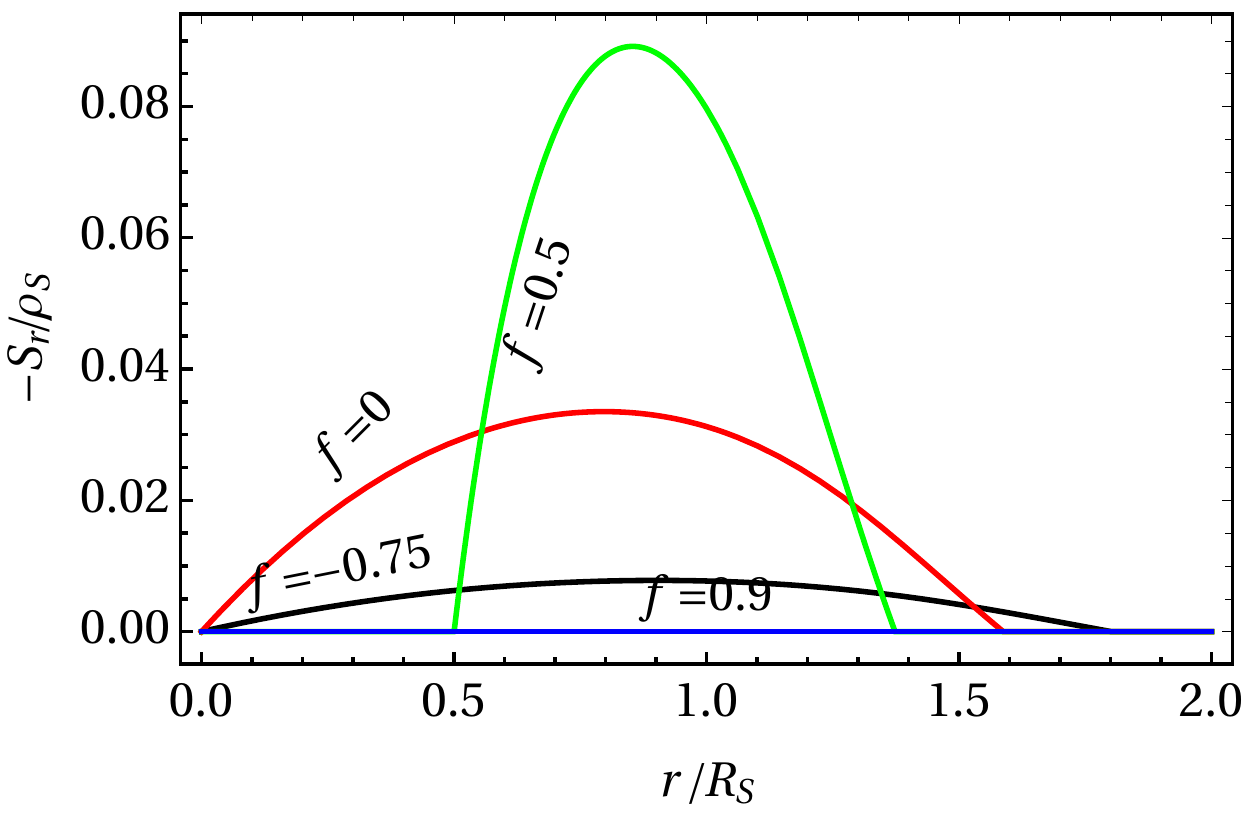}
    \caption{The tangential pressure $p_T$, the anisotropy $\Delta$, and the energy flow term $-S_r$ at the same stages of collapse as shown in Figs. \ref{rainbow den} and \ref{rainbow pr}. $\Delta$ and $p_T$ are large and positive in the inversion zone. $\Delta$ and $S_r$ are zero at $r=0$ and inside the dark energy core. $S_r$ is nonzero in the normal zone only. We plot $-S_r$ because it is the actual term in the stress energy tensor.}
    \label{Srainbow}
\end{figure}

\begin{figure}[H]
    \centering
   \begin{minipage}{0.8 \linewidth}
    \includegraphics[width=6.8cm]{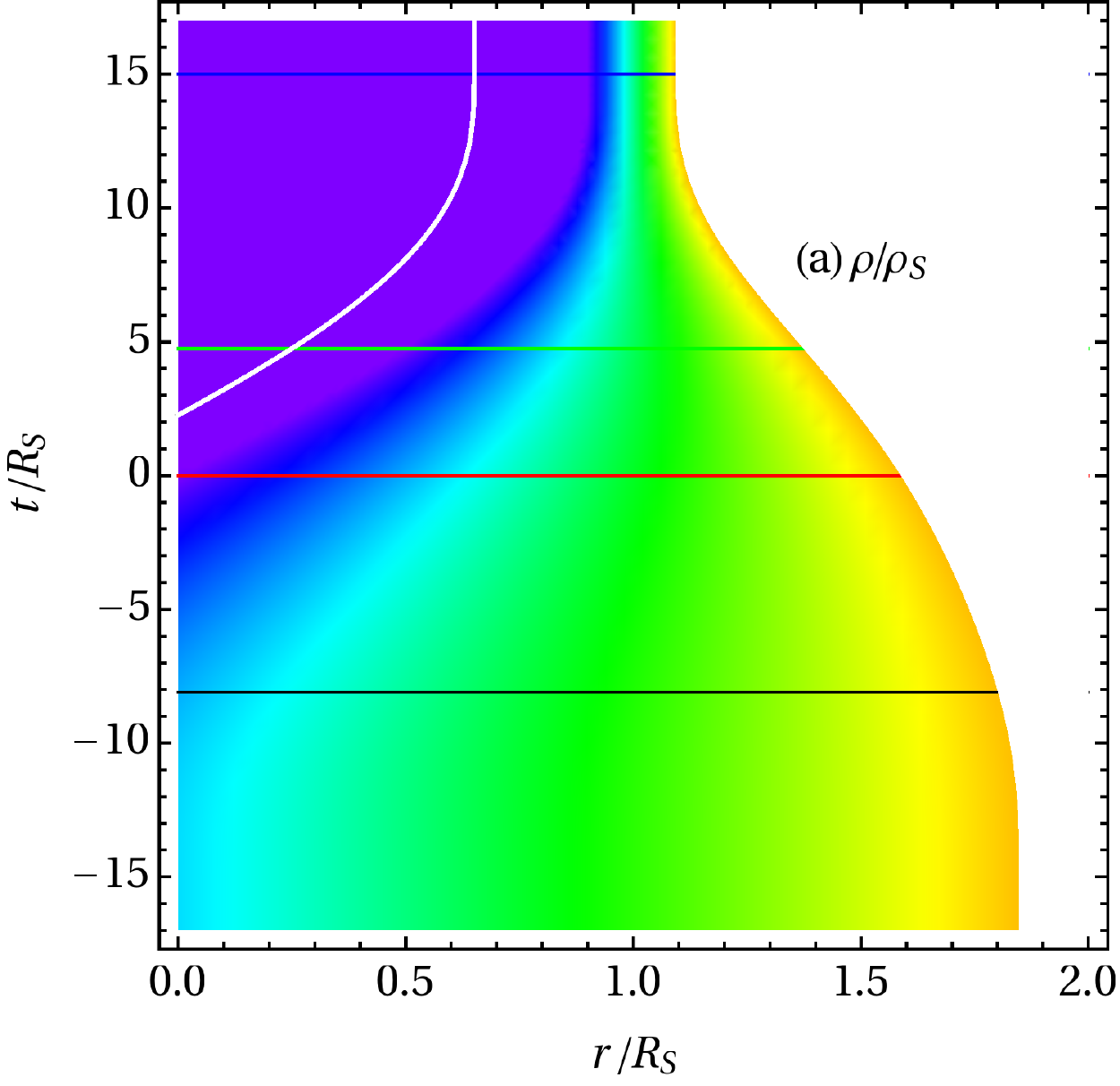}
    \includegraphics[width=6.8cm]{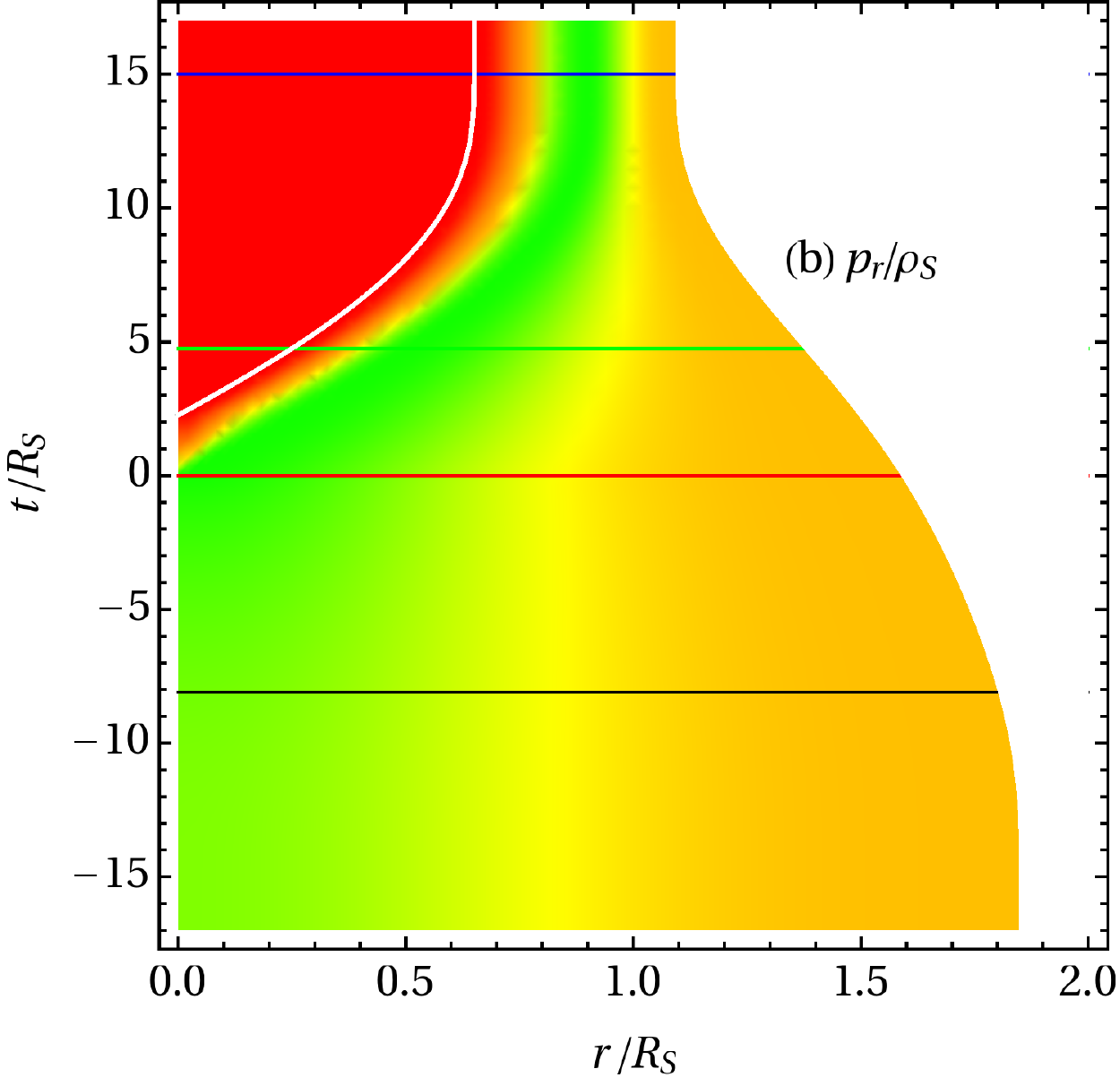}
     \includegraphics[width=6.8cm]{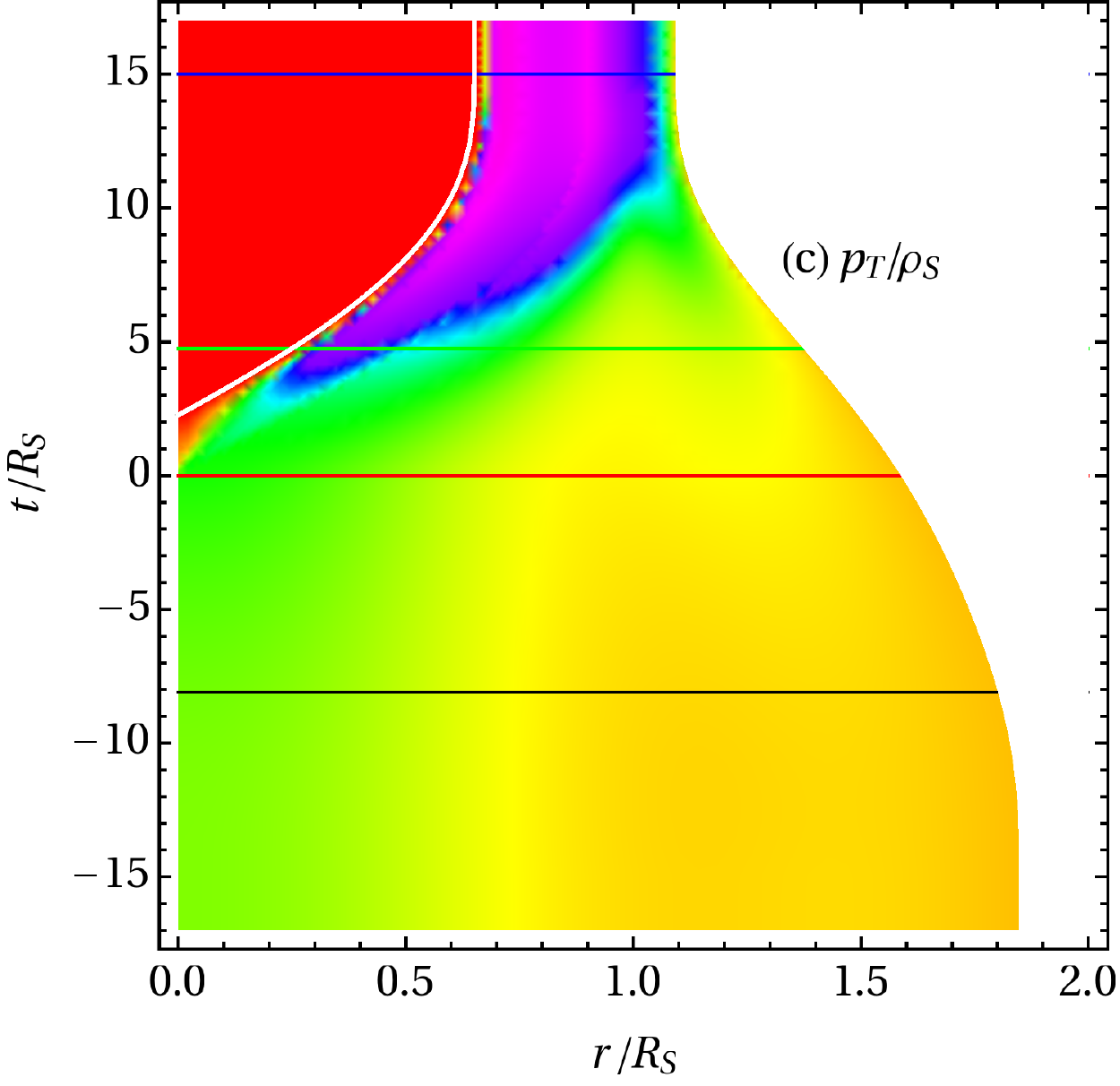}
    \includegraphics[width=6.8cm]{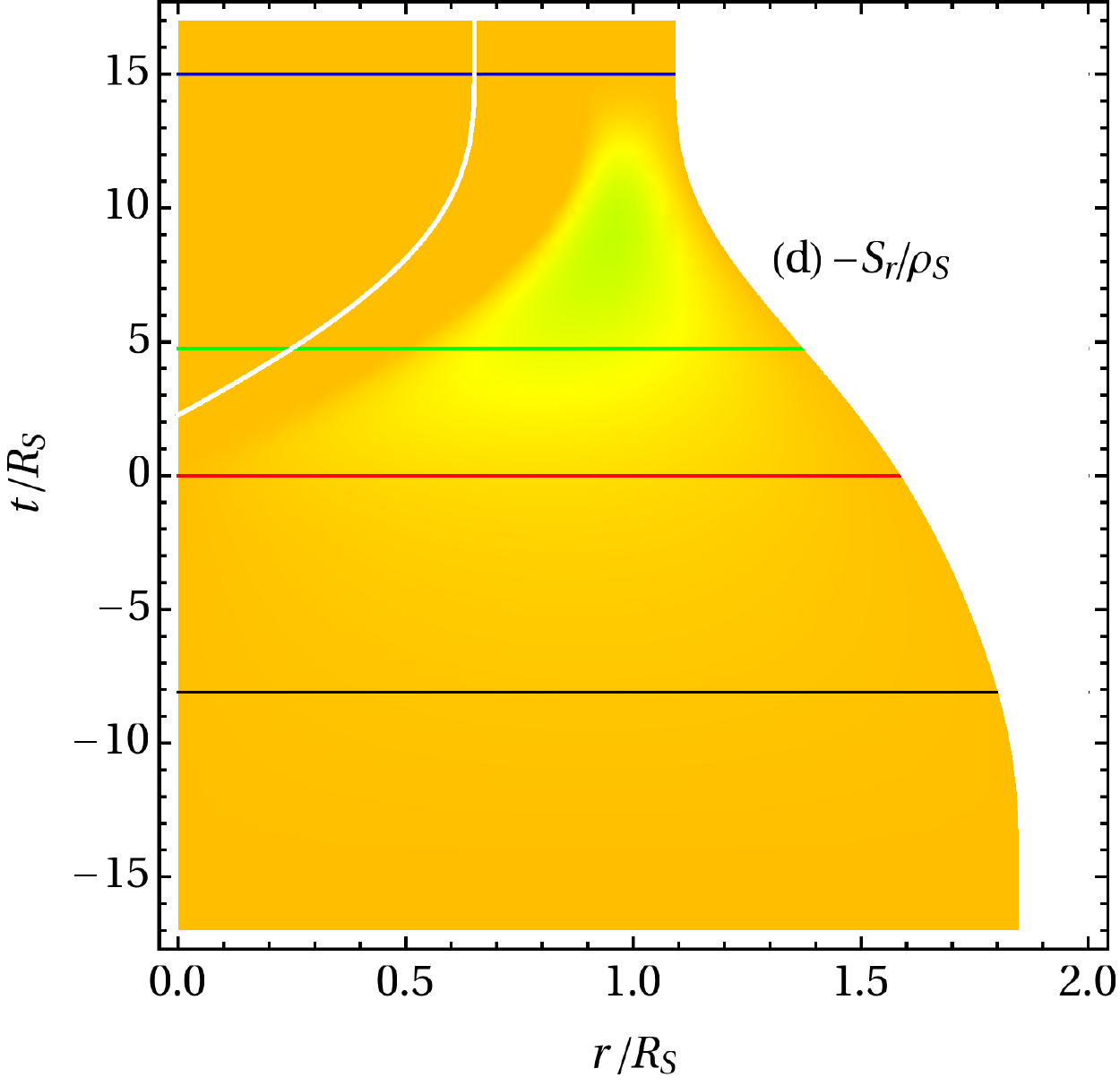}
    \end{minipage}
    \begin{minipage}{0.1 \linewidth}
    \includegraphics[width=1.5cm]{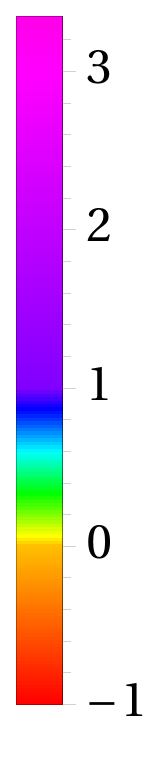}
    \end{minipage}
    \caption{ Plots of the $T_{\mu \nu}$ functions as functions of radius $r$ and time $t$: energy density $\rho$, radial pressure $p_r$, tangential pressure $p_T$, and energy flow/momentum density $-S_r$. The white region on the right is vacuum exterior. The white line on the top left delineates the dark energy core and the colored horizontal lines correspond to the profiles shown in Figs. \ref{rainbow den}, \ref{rainbow pr}, and \ref{Srainbow}. For $|t|>t_C/2$ the configurations are static. }
    \label{fig:my_label}
\end{figure}

 With the specified $f(t)$, we may calculate the evolution of $p_T$, $\Delta$, and $S_r$, which is displayed in Fig. \ref{Srainbow}. The tangential pressure $p_T$ and anisotropy $\Delta$ become large and positive in the inversion zone. The anisotropy $\Delta$ is zero for all times at $r=0$ and is zero inside the dark energy core. The energy flow $S_r$ is confined to the normal zone.
In Fig. \ref{fig:my_label} we show the evolution of the matter functions $\rho$, $p_r$, $p_T$, and $S_r$ over a wide range of $r$ and $t$. The white region on the right is the exterior of the object. The white line on the top left delineates the dark energy core. The colored horizontal lines correspond to the profiles shown in Figs. \ref{rainbow den}, \ref{rainbow pr}, and \ref{Srainbow}. One can see the formation and spread of the dark energy core in the $p_r$ and $p_T$ panels (red area) and the density plateau in the $\rho$ panel (purple area). Also, $p_r$ and $p_T$ are similar, showing low anisotropy, for a precursor $t<0$, and they become distinct, showing high anisotropy, in the transition and dark energy star stages. The energy flow $S_r$ is, in general, smaller than the other matter functions, is zero in the density plateau, and goes to zero at large times as the collapse starts and stops. We remark that we have no infinitesimally thin shells of matter because the functions $\rho$, $p_r$ and $m$ have sufficient continuity class that we never take a derivative of a discontinuity and get a Dirac delta function. The function $\rho(t,r)$ is $C^0$ in $r$ and $t$, $p_r(t,r)$ is $C^1$ in $r$ and $C_0$ in $t$, and $m(t,r)$ is $C^1$ in $r$ and $t$.

\begin{figure}[H]
    \centering
    \includegraphics[width=7cm]{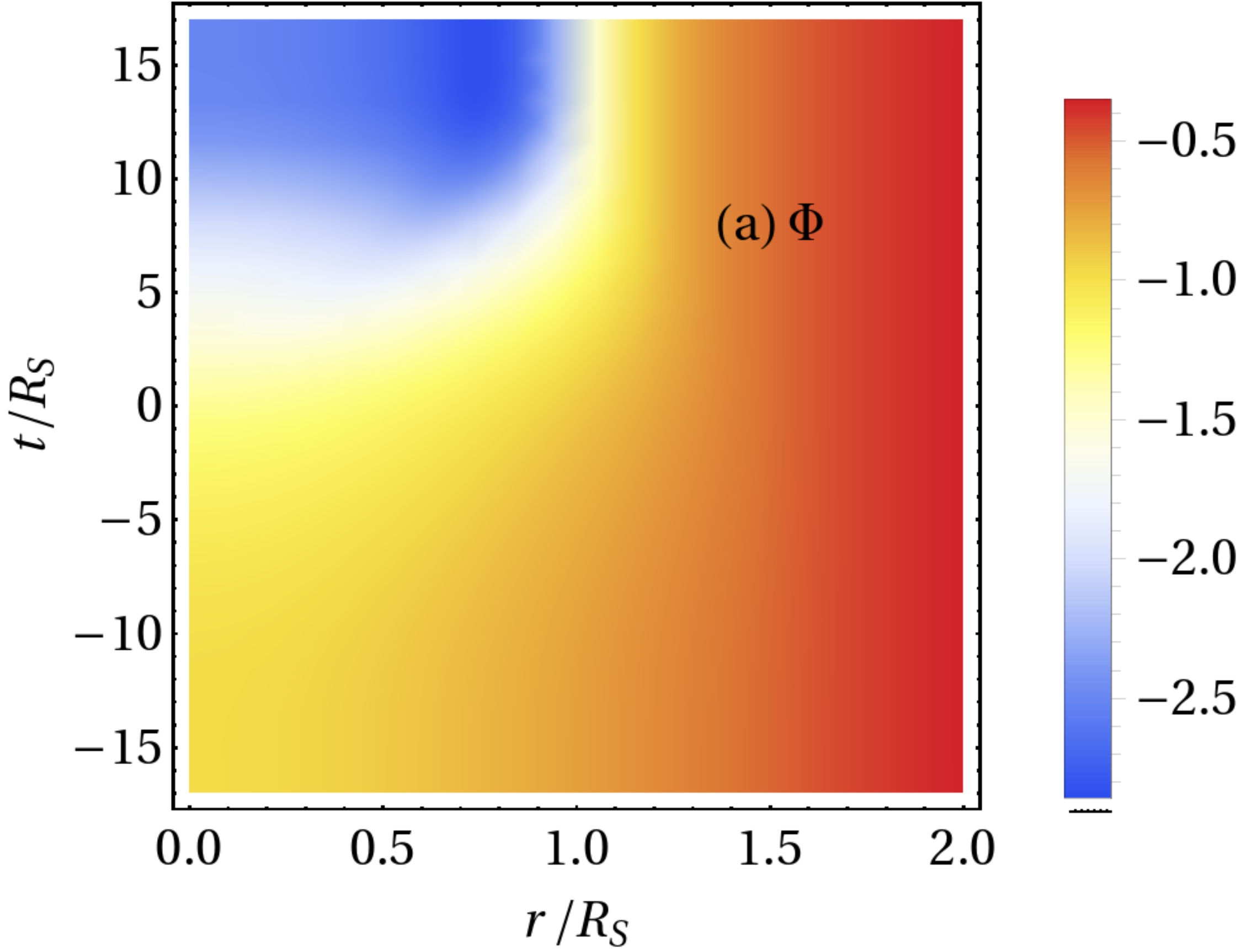}
    \includegraphics[width=6.8cm]{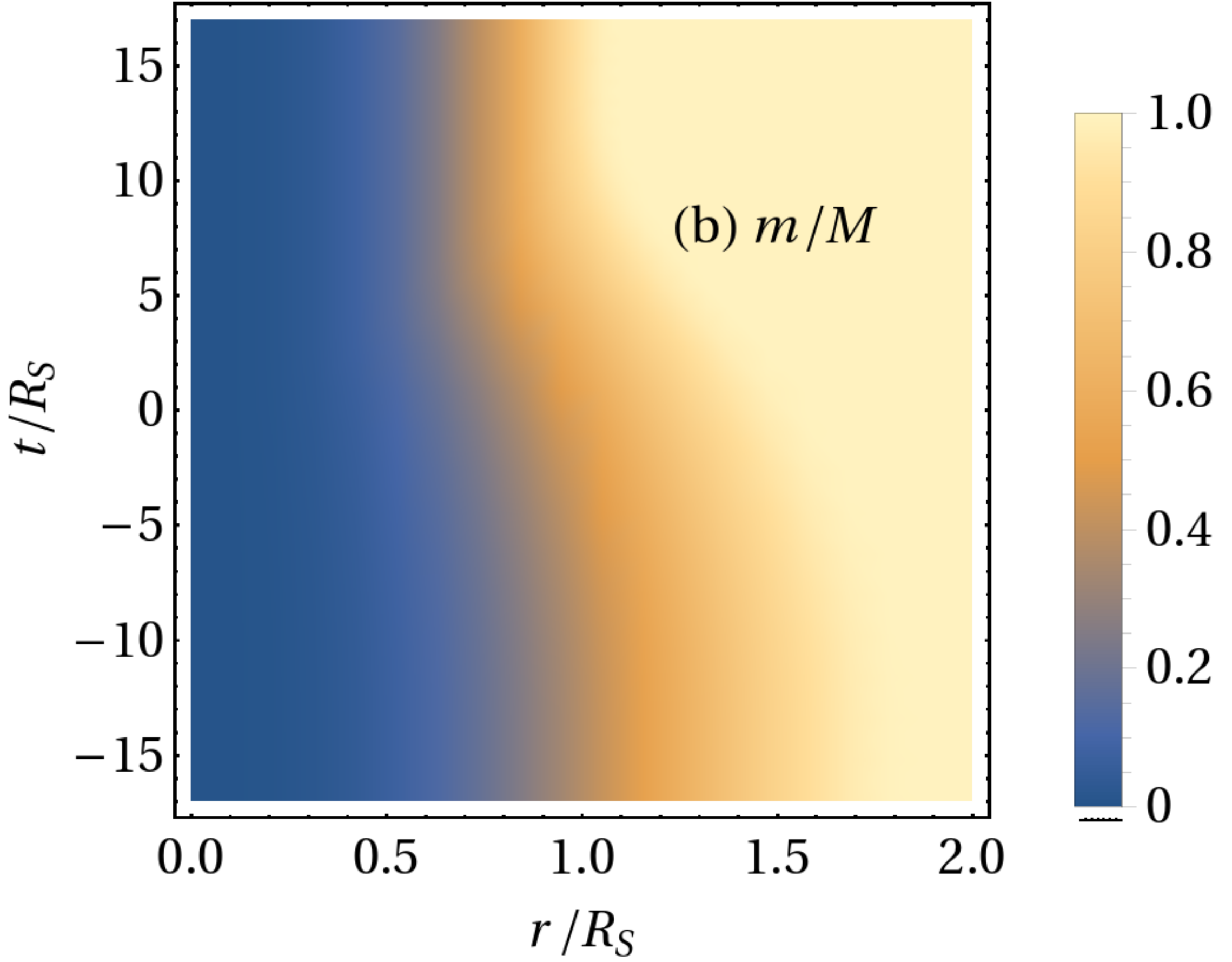}
    \caption{ Plots of the metric functions $\Phi(t,r)$ and $m(t,r)$. In the absence of singularities, $m(t,r)$ is constrained to be 0 at $r=0$ and $M$ in the vacuum exterior. $\Phi(t,r)$ is constrained to be 0 at $r=\infty$ and shows a minimum at $r\ne0$ after the formation of a dark energy core. In our case, $m$ is a simple piecewise polynomial function. $\Phi$ has analytic contributions from the dark energy core and vacuum exterior, but the contributions from the normal and inversion zones are not simple and are evaluated numerically. }
    \label{metricfunc}
\end{figure}

For completeness, we display plots of the metric functions $m(t,r)$ and $\Phi(t,r)$ in Fig. \ref{metricfunc}. The $m(t,r)$ function can be expressed analytically due to the simplicity of $\rho$ and the integral in Eq.~(\ref{eq:m(t,r)}). The contributions to the integral for $\Phi(t,r)$ [see Eq.~(\ref{eq:Phi(t,r)})] are analytic in the Schwarzschild vacuum and dark energy core, but the integral is evaluated numerically in the inversion zone and normal zone. In the absence of singularities $m(t,0)=0$ and $m(t,r)=M$ for $r>R(t)$. Also, $\Phi(t,\infty)=0$, and the minimum in $r$ of $\Phi$ is at $r=0$ for the precursor but at some $r\ne0$ for the dark energy star.
\section{Detailed Weak Energy Condition examination}
In this section, we examine the weak energy condition in detail, and we find that the example pileup model we defined does in fact satisfy the weak (and therefore null) energy condition.
\subsection{Dark Energy Core}
Within the plateau, $\partial m/\partial t=0$ and therefore $S_r=0$ and the energy condition inequalities are given in Eq.~(\ref{eq:wecstat}). The dark energy equation of state $p_r=p_T=-\rho$ satisfies these inequalities trivially.
\subsection{Inversion zone}
The inversion zone is within the plateau for our example so the energy conditions are still the static ones from Eq.~(\ref{eq:wecstat}). The first two are satisfied automatically by the construction of $\rho(t,r)$, $ p_r(t,r)$. The third may be shown to be true in the following way. We may write
\begin{equation}
   \rho+p_T= \rho+p_r+\Delta=\rho+p_r+\frac{r}{2}\frac{\partial p_r}{\partial r}+\frac{G \left( m+4\pi r^3 p_r \right) \left( \rho+p_r \right)}{2 r \left( 1 - \frac{2 G m}{r} \right) }.
\end{equation}
Since within the plateau region  $\partial p_r/\partial r \ge 0$, the following inequality is implied:
\begin{equation}
   \rho+p_T \ge (\rho+p_r)\Bigg[1+\frac{G \left( m+4\pi r^3 p_r \right) }{2 r \left( 1 - \frac{2 G m}{r} \right) }\Bigg].
\end{equation}
Since $\rho+p_r\ge0$ and $ 1-2 G m/r\le 1$, one has
\begin{equation}
   (\rho+p_r)\Bigg[1+\frac{G \left( m+4\pi r^3 p_r \right) }{2 r \left( 1 - \frac{2 G m}{r} \right) }\Bigg]\ge  (\rho+p_r)\Bigg[1+\frac{G \left( m+4\pi r^3 p_r \right) }{2 r}\Bigg].
\end{equation}
Within the plateau region $m=\frac{4}{3}\pi r^3 \rho_S$, and $p_r\ge-\rho_s$, therefore $m+4\pi r^3 p_r\ge-\frac{8\pi}{3}r^3 \rho_S$, meaning
\begin{equation}
    (\rho+p_r)\Bigg[1+\frac{G \left( m+4\pi r^3 p_r \right) }{2 r}\Bigg]\ge(\rho+p_r)\left(1-\frac{4\pi G \rho_S r^2}{3}\right).
\end{equation}
Using $r\le R_S$ in the plateau region we obtain
\begin{equation}
   (\rho+p_r)\left(1-\frac{4\pi G \rho_S r^2}{3}\right)\ge(\rho+p_r)\left(1-\frac{4\pi G \rho_S R_S^2}{3}\right)=(\rho+p_r)\left(1-\frac{G M}{R_S}\right)=(\rho+p_r)\left(1-\frac{1}{2}\right)=\frac{\rho+p_r}{2}.
\end{equation}
Again using the fact that $\rho+p_r\ge0$ by construction, we may conclude
\begin{equation}
  \rho+p_T \ge \frac{\rho+p_r}{2} \ge0 .
\end{equation}
The last weak energy condition inequality is therefore satisfied.
\subsection{Normal region}
In this region there is momentum present, so we need to examine the full energy conditions Eqs.~(\ref{eq:wec3})-(\ref{eq:wec4}). In certain cases however, the applicability conditions allow us to make simplifications. For the cases in Eqs.~(\ref{eq:wec3})-(\ref{eq:wec4}), we proceed as follows.
\subsubsection{Equation~(\ref{eq:wec3}) }
Because of the condition of applicability of Eq.~(\ref{eq:wec3}), it follows that $\rho-\frac{S_r^2}{p_r}\ge \rho-p_r$. Because of the value we set for $a$ in our example $\rho\ge p_r$, we have 
\begin{equation}
    \rho-\frac{S_r^2}{p_r}\ge \rho-p_r\ge0.
\end{equation}
The inequality from Eq.~(\ref{eq:wec3}) is therefore satisfied when applicable.
\subsubsection{Equation~(\ref{eq:wec2}) }
Note that within the normal region $\rho$ and $p_r$ are nonnegative, but the inequality from Eq.~(\ref{eq:wec2}) can still be violated if $S_r$ is too high. 

\begin{figure}[H]
    \centering
    \includegraphics[width=8cm]{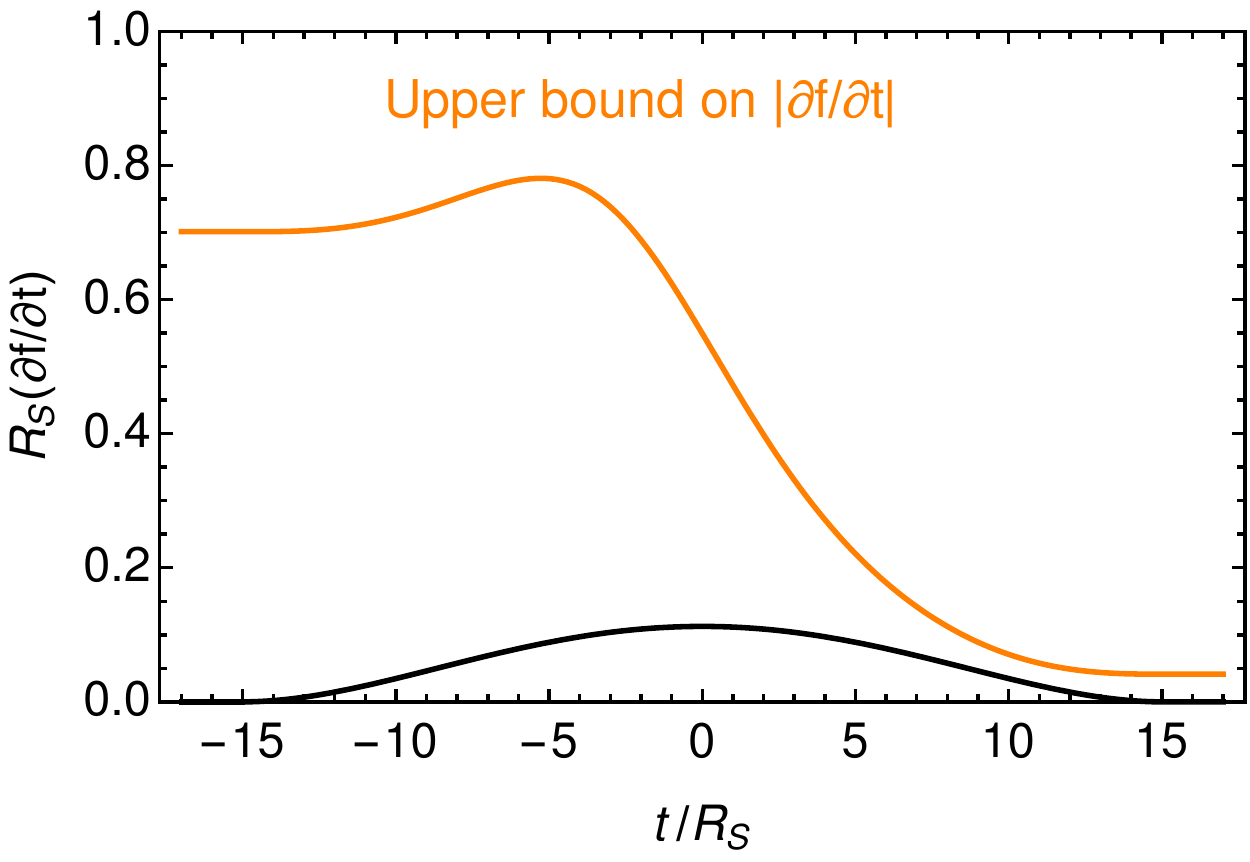}
    \caption{Check of the WEC (\ref{eq:wec2}) in the normal zone. The condition in Eq.~(\ref{superf}) imposes $\partial f/\partial t$ to be below the orange line. The black line is $\partial f/\partial t$ for our choice of $f$ Eq.~(\ref{ftau}). Thus we see that our choice satisfies the WEC (\ref{eq:wec2}) in the normal zone.}
    \label{wec1proof}
\end{figure}

 We may reexpress $\rho+p_r-2|S_r|\ge0$ as
\begin{equation}
  \Big|\frac{\partial m}{\partial t}\Big| \le  4 \pi  r^2 e^{\Phi}\sqrt{1-\frac{2 G m}{r}}\frac{\rho+p_r}{2}\text{ for all $r$}.
\end{equation}
 We may use the chain rule $\frac{\partial m}{\partial t}=\frac{\partial m}{\partial f}\frac{\partial f}{\partial t}$, and the fact that $\frac{\partial f}{\partial t}$ is independent of $r$, to find an equivalent condition on $f$ rather than $m$.
\begin{equation}
     \Big|\frac{\partial f}{\partial t}\Big| \le \min_{r\text{ } \epsilon\text{ normal zone}}\left( 4 \pi  r^2 e^{\Phi}\sqrt{1-\frac{2 G m}{r}}\frac{\rho+p_r}{2|\frac{\partial m}{\partial f}|}\right),
     \label{superf}
\end{equation}
and Eq.~(\ref{superf}) is a constraint on $\partial f/\partial t$, which we show in Figure \ref{wec1proof} together with our choice of $f(t)$. We see from the figure that the inequality (\ref{superf}) is clearly satisfied. Therefore a rest frame for the matter exists and the condition from Eq.~(\ref{eq:wec2}) is satisfied in the normal region. 

\subsubsection{Equation~(\ref{eq:wec4})}
Because of the applicability condition $\Delta \le-|S_r|$, the inequality from Eq.~(\ref{eq:wec4}) is implied by $\rho+p_r+2\Delta\ge0$ where applicable. In terms of the anisotropy force, $F_\Delta=2\Delta/r$ one has $F_\Delta\ge(\rho+p_r)/r$. Since $F_G\le0$ in the normal zone, one may write $(\rho+p_r)/(r F_G)\ge F_\Delta/F_G$. We can then simplify the expression on the left with the form of $F_G$ and rewrite as the following:
\begin{equation}
    \frac{r-2Gm}{G(m+4\pi r^3 p_r)}\ge\frac{F_\Delta}{F_G}.
    \label{sufficient}
\end{equation}
The interpretation of this sufficient condition is as follows: WEC (\ref{eq:wec4}) is satisfied within this region, if the anisotropy force is not pulling ``in" too strongly.

\begin{figure}[H]
    \centering
    \includegraphics[width=8cm]{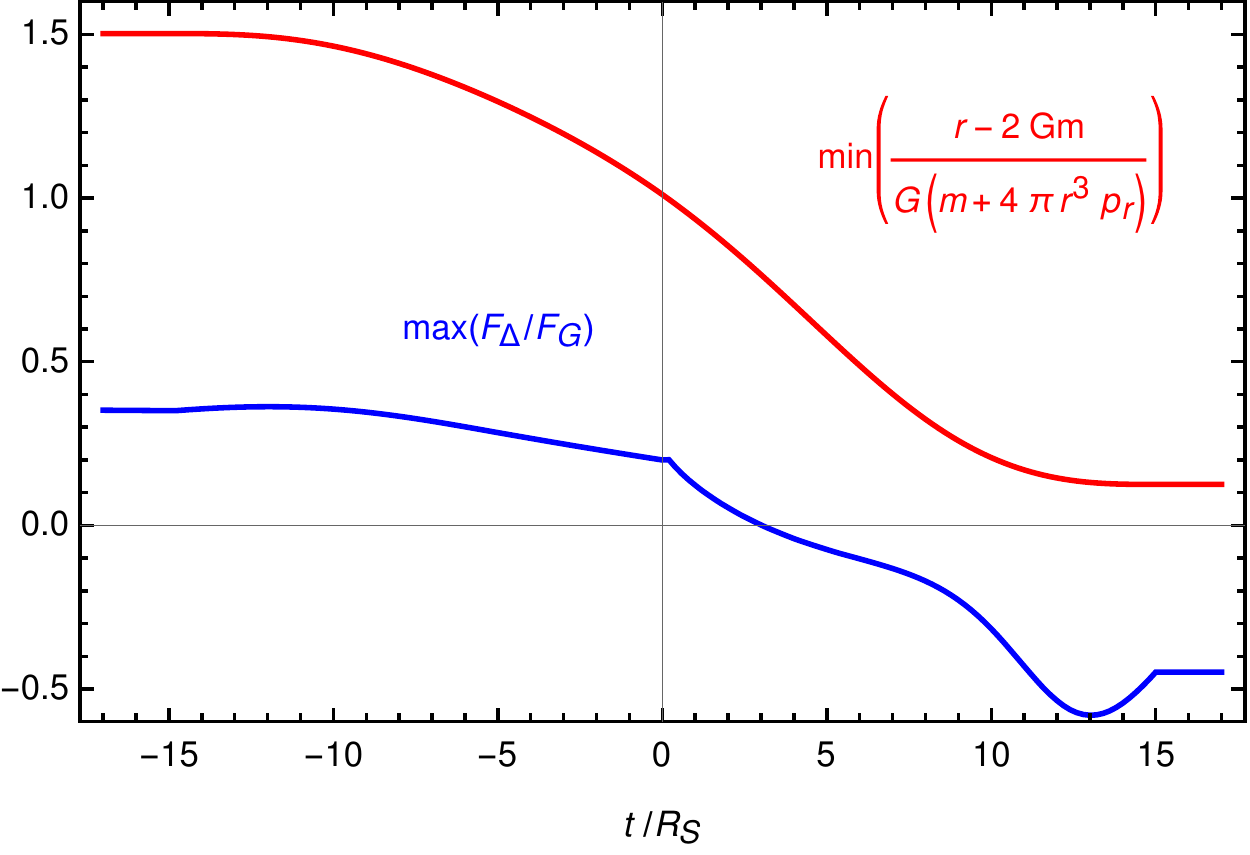}
    \caption{The maximum of the force ratio $F_\Delta/F_G$ (blue) is below the minimum of the left-hand side of Eq.~(\ref{sufficient}) (red) for all $f$. Therefore, the third WEC inequality (\ref{eq:wec4}) is satisfied in the normal region.}
    \label{wec3forces}
\end{figure}

In order to examine this, we look within the region for the maximum in $r$ of the ratio of the forces $F_\Delta/F_G$ and compare it to the minimum in $r$ of the left side of Eq.~(\ref{sufficient}). We graph both in Fig. \ref{wec3forces} and conclude the inequality (\ref{sufficient}) is always satisfied. 

\subsection{Minimum \boldmath$T_{\mu \nu}k^\mu k^\nu$ Summary}
\begin{figure}[H]
    \centering
    \includegraphics[width=8cm]{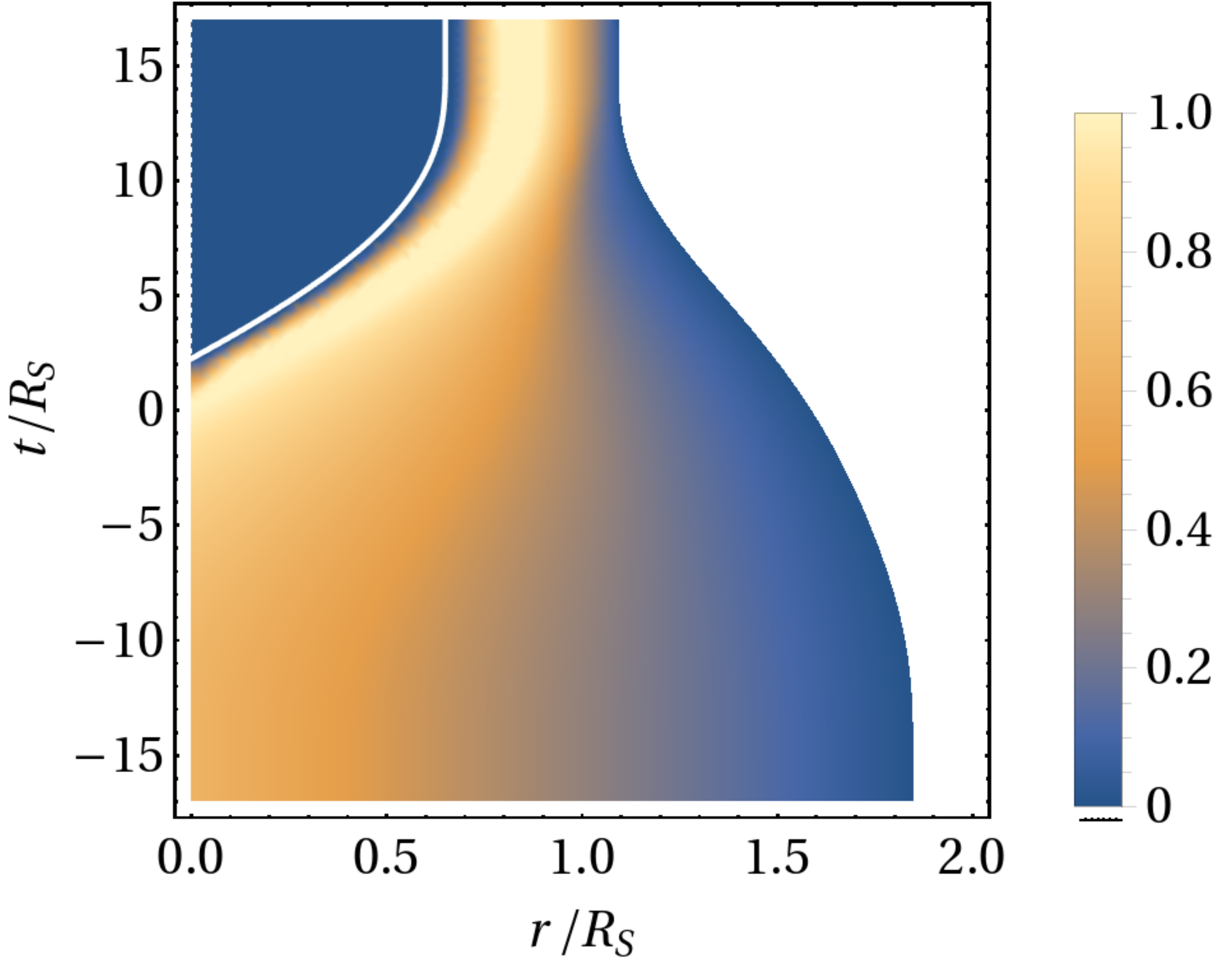}
    \caption{Minimum of $T_{\mu \nu}k^\mu k^\nu/\rho_S$ plotted at points in $r,t$. The dark energy core is the region on the top left bounded by the white curve. The white region on the right is the exterior of the collapsing object. Note the formation and spread of the dark energy core where $T_{\mu \nu}k^\mu k^\nu=0$. Also note that $T_{\mu \nu}k^\mu k^\nu$ is non-negative at all points, meaning the weak energy condition is satisfied at all times and positions during collapse.}
    \label{contour}
\end{figure}

As a summary, Fig. \ref{contour} is a graph of the minimum of $T_{\mu \nu}k^\mu k^\nu$ over position and time. We see that this minimum is non-negative at all points, being zero in the dark energy core and vacuum exterior and positive in the normal and inversion zones.
\section{Conclusion}
We have presented a model for dark energy stars that describes the collapse of a spherical object from an initial state of positive pressure to a final state with negative pressure (equation of state $p=-\rho$) inside a finite radius core. Our model contains no spacetime or coordinate singularities, no event horizons, and it satisfies the weak and null energy conditions. In the static case, dark energy stars offer an ultracompact object with no singularities or event horizons. Our work shows that dynamical formation can still satisfy these criteria.

The strong energy condition is violated due to the $p=-\rho$ region. The dominant energy condition is violated in our particular example by a large positive $p_T$, although it appears that the DEC may be satisfied by less compact dark energy stars that require less anisotropy (i.e., $f_\infty\le0.43$). In any case, dominant energy condition violations due to high tangential pressure are recognized as common in anisotropic gravastar systems \cite{eosgravastar}.  

We feel it is worth mentioning explicitly why the various singularity theorems (see Ref. \cite{Hawking:1973uf}) do not apply to our system. Penrose's 1965 singularity theorem, originally published in Ref. \cite{PhysRevLett.14.57}, states that singularities will result from gravitational collapse when the curvature condition $R_{\mu \nu}k^\mu k^\nu \ge0$ for null vectors $k^\mu$ (which is equivalent to the null energy condition by Einstein's equation \cite{Curiel2017}) is satisfied, a global Cauchy hypersurface exists, and a closed trapped surface exists. No closed trapped surface forms in our system, so this theorem does not apply. Two additional theorems by Hawking from 1967 and one by Hawking and Penrose in 1970 are more general in that criteria other than trapped surfaces can imply singularities. However, these theorems require the curvature condition $R_{\mu \nu}k^\mu k^\nu \ge0$ for all timelike vectors $k^\mu$, which is equivalent to the strong energy condition \cite{Hawking:1973uf,Curiel2017} and is violated in our system because it contains $p=-\rho$ dark energy. Buchdahl's theorem \cite{PhysRev.116.1027} is inapplicable both because the trace of the stress energy tensor in our system may be negative and the pressure may be anisotropic.

 In spherical symmetric systems, once a dark energy core is formed its density cannot change without violating the weak energy condition. As such, we have introduced the idea of pileup models, and we have shown an example of a pileup model where the weak energy condition is satisfied and no event horizon or singularity is formed. By reexpressing the spherically symmetric Einstein field equations, we have defined our model in terms of a density function and radial pressure function. Defining the system in terms of matter functions is conducive to an easy evaluation of the energy conditions.

Alternatively, one could have specified some equation of state, or perhaps multiple equations of state, for anisotropic matter, then one could have used Eq.~(\ref{Eq:Force}) as a force equation and Eq.~(\ref{eq:Continutiy}) as a continuity equation to solve for the time evolution. Finding equations of state and initial conditions that result in formation of dark energy stars without singularities or event horizons while maintaining the WEC is an area that requires further research.

\section{Acknowledgements}
 P.G. thanks Emil Mottola for an intriguing conversation that rekindled his interest in gravastars, and Stefano Ansoldi, Antonio De Felice, Shinji Mukohyama, Fumihiro Takayama, and Takahiro Tanaka for helpful discussions on the topic of this paper. P.G. also thanks the Yukawa Institute for Theoretical Physics at Kyoto University where part of this work was carried out. This work has been partially supported by NSF Award PHY-1720282 at the University of Utah.

\medskip

\bibliographystyle{apsrev4-1}
\bibliography{X.bib}

\appendix

\end{document}